\documentclass[11pt]{article}
\usepackage[left=1.2in,right=1.2in,top=1.2in,bottom=1.3in]{geometry}
\usepackage{amsmath}
\usepackage{amsfonts}
\usepackage{bm}
\usepackage{graphicx}
\usepackage{subfigure}
\usepackage{tikz-cd}
\usepackage[affil-it]{authblk}
\usepackage{setspace}

\usepackage{titlesec}
\titleformat*{\section}{\Large\bfseries}
\titleformat*{\subsection}{\large\bfseries}
\titleformat*{\subsubsection}{\large\itshape}

\makeatletter
\def\@maketitle{%
  \newpage
  \begin{flushright}\large EFI 14-40\end{flushright}
  \null
  \vskip 4em%
  \begin{center}%
  \let \footnote \thanks
    {\huge \@title \par}%
    \vskip 1.6cm%
    {\large
      \lineskip .5em%
      \begin{tabular}[t]{c}%
        \@author
      \end{tabular}\par}%
  \end{center}%
  \par
  }
\makeatother

\usepackage{abstract}

\usepackage{feynmp}
\usepackage[pdfstartview={FitH},pdfpagelayout=OneColumn, colorlinks=true,pdfhighlight=/I,bookmarksnumbered=true,bookmarksopen=true,bookmarksopenlevel=1, debug=false,linkcolor=blue,citecolor=blue,urlcolor=blue,pdfpagemode=UseOutlines,pdffitwindow=false]{hyperref}

\numberwithin{equation}{section}

\newcommand{\overbar}[1]{\mkern 1.5mu\overline{\mkern-1.5mu#1\mkern-1.5mu}\mkern 1.5mu}

\def\var{\phipert}
\def\varn{\overbar{\var}}
\def\bq{\mathbf{q}}
\def\vacl{\bigl\langle0\bigr|}
\def\vacr{\bigl|0\bigr\rangle}
\def\Ivacl{\bigl\langle\Omega\bigr|}
\def\Ivacr{\bigl|\Omega\bigr\rangle}
\def\bp{\mathbf{p}}
\def\phin{\overbar{\phipert}}

\def\sfg{_{\rm s.f.}}
\def\cfg{_{\rm c.i.}}
\def\2d{$1+1$ dimensions}
\def\4d{$3+1$ dimensions}
\def\ct{\tau} 
\def\pht{t} 

\def\Xb{\tilde X}

\def\bdm{\begin{displaymath}}
\def\edm{\end{displaymath}}

\def\lr{\leftrightarrow}

\newcommand{\dX}{{\mathsf X}}
\def\mk{k}
\def\bx{\mathbf{x}}

\def\bk{\mathbf{k}}

\def\({\left(}
\def\){\right)}
\def\[{\left[}
\def\]{\right]}

\def\barray{\begin{array}}
\def\earray{\end{array}}
\def\be{\begin{equation}}
\def\ee{\end{equation}}
\def\ben{\begin{equation} \nonumber}
\def\een{\end{equation}}
\def\ban{\begin{eqnarray*}}
\def\ean{\end{eqnarray*}}
\def\bea{\begin{eqnarray}}
\def\eea{\end{eqnarray}}
\def\eal{\end{align}}
\def\bal{\begin{align}}

\def\({\left(}
\def\){\right)}
\def\half{{1\over2}}

\def\One{{\hbox{ 1\kern-.8mm l}}}

\def\px{\partial_{x}}
\def\pt{\partial_{t}}

\def\HH{{\mathcal{H}}}

\def\SS{{\mathcal{S}}}

\def\VV{{\mathcal{V}}}

\def\ghat{\hat g}

\def\R{R}

\def\grav{{\rm grav}}

\def\inflaton{X}		
\def\infb{\tilde{X}}	
\def\infpert{\bm x}	
\def\infsub{{\scriptscriptstyle X}}	
\def\dX4d{{\delta \!X}}	

\def\Laux{\chi}			
\def\Lauxb{\tilde\chi}		
\def\Lauxpert{{\bm \chi}}	

\def\phican{\phi}		
\def\phib{{\tilde{\phican}}}	
\def\phipert{\bm\varphi}	

\def\ntpert{{{\bm n}^\ct}}	
\def\nxpert{{{\bm n}^x}}	

\def\dS{de~Sitter }

\def\infpertd{\infpert'}

\newcommand{\cH}{\mathcal{H}}
\newcommand{\cP}{\mathcal{P}}

\newcommand{\pd}[2]{\frac{\partial #1}{\partial #2}}

\newcommand{\fd}[2]{\frac{\delta #1}{\delta #2}}
\newcommand{\fdd}[2]{\frac{\delta^{2} #1}{\delta #2}}

\title{Primordial fluctuations in extended Liouville theory}
\author{Wynton E. Moore\thanks{wyntonmoore@uchicago.edu}}
\affil{Enrico Fermi Institute and Department of Physics\\ University of Chicago, Chicago, IL 60637, USA}
\date{}

\begin{document}
\onehalfspacing

\maketitle
\thispagestyle{empty}

\begin{abstract}\large
Liouville gravity can be used to precisely model features of 3+1 dimensional cosmology in a simplified 1+1d setting. We study primordial fluctuations in a generally covariant extension of Liouville theory, in the context of single field inflation. The scale invariant spectrum of scalar curvature perturbations is exhibited, and their three-point correlation function is computed in the slow roll approximation. We recover Maldacena's consistency relation for the three-point function, which in this context depends on a global shift symmetry of extended Liouville theory.
\end{abstract}
\clearpage

\setcounter{page}{1}
\section{Introduction}\label{introsec}
One of the cornerstones of modern cosmology is the framework of cosmological perturbation theory, which allows detailed predictions to be derived from fundamental models of the early universe. The theory of gauge-invariant cosmological perturbations was pioneered in \cite{Bardeen:1980kt}, and the resulting power spectrum of curvature perturbations in single field inflation was first computed in \cite{Sasaki:1986hm,Mukhanov:1988jd}; see \cite{Mukhanov:1990me,Baumann:2009ds} for reviews. The three-point function (bispectrum) in single field inflation was computed in  \cite{Maldacena:2002vr}, a calculation which has been extended to many other models of inflation. More recently there has been interest in the consequences of Ward identities for cosmological correlation functions, arising from spatial diffeomorphisms and conformal symmetry \cite{Creminelli:2012qr,Hinterbichler:2013dpa,Berezhiani:2013ewa,Pimentel:2013gza}.

Theoretical interest of a different kind is attached to gravity in two spacetime dimensions, where Newton's constant is dimensionless and the gravitational interaction is renormalizable. The canonical theory of gravity in 2d is Liouville theory \cite{Polyakov:1981rd}, which describes the conformal factor of the 2d metric in conformal gauge. Liouville theory has attracted sustained interest as a toy model of quantum gravity, and as an interacting conformal field theory whose classical solutions are known in full \cite{Braaten:1982fr,Braaten:1982yn,Teschner:2001rv,Polchinski:1989fn,Ginsparg:1993is} (although the status of timelike Liouville as a conformal field theory remains unclear \cite{McElgin:2007ak,Harlow:2011ny}). It also has relevance for cosmology, providing a description of two-dimensional (A)dS spacetimes, and giving rise to 2d Friedmann equations when coupled to conformally invariant matter \cite{DaCunha:2003fm}.

General covariance is a powerful organizing principle in the theory of cosmological perturbations \cite{Bardeen:1980kt}. With this connection in mind, a generally covariant extension of Liouville theory was recently introduced \cite{Martinec:2014uva}. At the classical level, this extended Liouville theory is a local version of Polyakov's covariant 2d gravity \cite{Polyakov:1987zb}. It features an auxiliary scalar field in addition to the 2d metric, making for a total of four scalar degrees of freedom --- including two Lagrange multipliers --- subject to a gauge symmetry consisting of two-component diffeomorphisms. This counting is identical to the scalar sector of perturbations in 4d Einstein gravity \cite{Bardeen:1980kt}, and indeed there is a precise map between perturbations in the two models.

The purpose of this paper is to continue the analysis of cosmological perturbations in extended Liouville theory coupled to a scalar inflaton, begun in \cite{Martinec:2014uva}. We begin in sections \ref{Liouvillesec} and \ref{quadsec} with a review of the basic features of extended Liouville theory, its perturbative degrees of freedom and their correspondence with the scalar sector in 4d. We construct the systematic slow roll expansion and give examples of spatially homogeneous slow roll backgrounds. The power spectra of scalar curvature perturbations in 2d and 4d are derived in a unified manner, making transparent the characteristic scale invariance. We also confirm the freezing out of scalar curvature perturbations on superhorizon scales, to all orders in perturbation theory.

In section \ref{cubicsec} the cubic fluctuation action is derived in spatially flat gauge and constant inflaton gauge. Following the method of \cite{Maldacena:2002vr}, we confirm the suppression of the cubic action by two powers of the slow roll parameter $\epsilon$. This allows us to evaluate, in section \ref{3ptsec}, the three-point function of scalar curvature perturbations in the slow roll approximation, by working at tree level and using the cubic vertex only. The resulting non-Gaussianity has a local shape function analogous to that generated by single field inflation in 4d. Kinematic simplifications in 1+1d allow us to condense the shape function into a very compact form.

Finally, in section \ref{consistencysec} we discuss consistency relations for the three-point function. It turns out that there is no room in 1+1d for residual diffeomorphisms preserving the gauge choice and the physicality of metric fluctuations. However, we find that spatial dilations are revived by a global shift symmetry of extended Liouville theory, leading to the familiar consistency relation of Maldacena. We show how this shift symmetry can be interpreted as a global Weyl symmetry of the theory in cosmological gauges, which is related to the local Weyl symmetry of regular Liouville theory. There are no higher order consistency relations for the three-point function in 2d.

Emphasis throughout is on the similarity of our methods and results to those familiar in the scalar sector in 4d, lending support to the use of extended Liouville theory as a theoretical laboratory. This approach has already yielded new insights into non-perturbative effects of quantum gravity in inflation \cite{Martinec:2014uva}. It also provides a tool to study renormalized scalar perturbations in cosmological gauges, a topic which is the focus of ongoing work.

\section{Extended Liouville theory}\label{Liouvillesec}
Here we review the generally covariant extension of Liouville theory which was introduced in \cite{Martinec:2014uva}, discussing its basic features and interpretation in worldsheet string theory.

\subsection{Action, equations of motion, and constraints}
The action for the extended Liouville theory introduced in \cite{Martinec:2014uva} is\footnote{The action \eqref{Lcovpre} has been scaled by $4\pi$ relative to that in \cite{Martinec:2014uva}, to give the conventional normalization for the scalar curvature perturbation.}
\be
\label{Lcovpre}
\SS_\grav = \frac{2}{\gamma^2} \int \sqrt{-g}\left[  -(\nabla\Laux)^2 -  \R \Laux - {\Lambda} \right] 
\ee
Here $\Laux$ is an auxiliary scalar field, and $\gamma^{2}$ is a dimensionless parameter which plays the role of Newton's constant $G_{N}$. Precisely speaking, $2/\gamma^{2}$ plays the role of $m_{p}^{2}/2=1/16\pi G_{N}$, where $m_{p}$ is the Planck mass. All action integrals are over 1+1d spacetime, unless otherwise indicated. Let us see what sort of spacetimes are described by the theory \eqref{Lcovpre}. The equation of motion for $\Laux$ is
\begin{equation}\label{chieom}
0 = 2\nabla^{2}\Laux - R 
\end{equation}
The scalar curvature acts as a source for $\Laux$. The gravitational stress tensor\footnote{Strictly speaking gravity does not have a stress tensor, but by an abuse of language we use this terminology to denote the variation of the gravitational action with respect to the metric --- that is, the metric equations of motion.}
\begin{equation}
T_{ab}^{\rm grav} \equiv  \frac{2}{\sqrt{-g}} \fd{\SS_{\rm grav}}{g^{ab}}
\end{equation}
is equal to
\begin{equation}\label{Tgrav}
\gamma^{2} T_{ab}^{\rm grav} =  \half g_{ab}\[ (\nabla\Laux)^{2}+\Lambda \]
- \nabla_{a}\Laux\nabla_{b}\Laux
+ \nabla_{a}\nabla_{b}\Laux - g_{ab}\nabla^{2}\Laux
\end{equation}
The final two terms are ``improvement'' terms arising from the coupling to the Ricci scalar in \eqref{Lcovpre}. In 2d there is only one component of curvature, which completely determines the tensor structures
\begin{equation}\label{2dcurv}
R_{abcd} = \half R\( g_{ac}g_{bd}-g_{ad}g_{bc} \)~,\qquad 
R_{ab} = \half R\, g_{ab}
\end{equation}
These relations cause the Einstein tensor to vanish identically in 2d. In other words, the Einstein-Hilbert action is a topological invariant of the 2d manifold $M$, called the Euler characteristic $\chi_{\rm Euler}$ (not to be confused with the auxiliary field $\Laux$):
\begin{equation}\label{Euler}
\chi_{\rm Euler} (M) = \int_{M}\sqrt{-g}\,R
\end{equation}
In light of the vanishing Einstein tensor, the variation of the Einstein-Hilbert Lagrangian density with respect to the metric reduces to
\begin{equation}
\delta(\sqrt{-g}\, R) = \sqrt{-g}\,g^{ab}\delta R_{ab}
\end{equation}
which is a total derivative. In the Einstein-Hilbert action this would be discarded, but in the action \eqref{Lcovpre} it gives rise to the final two terms in the stress tensor \eqref{Tgrav}. The metric equations of motion state that all components of \eqref{Tgrav} must vanish. The trace component is
\begin{equation}\label{phieom}
0 = \nabla^{2}\Laux - \Lambda
\end{equation}
Clearly the solutions to \eqref{chieom} and \eqref{phieom} are metrics of constant curvature:
\begin{equation}\label{RLambda}
R[g]=2\Lambda
\end{equation}
This tells us everything about the curvature of the 1+1d manifold. The solutions are two-dimensional (anti-) \dS spacetimes. 

The remaining components of the metric equations of motion \eqref{Tgrav} are the Hamiltonian and momentum constraints, which enforce diffeomorphism invariance. The constraints are expressed most easily in terms of the conjugate momenta. Let us parametrize the metric in 1+1d as
\be
\label{metparam}
g_{ab}  = e^{2\phican}\left(  \begin{matrix} -N_\ct^2+N_x^2\quad & N_x \\ N_x & 1 \end{matrix}\right) 
\ee
where the lapse $N^{\ct}$ and shift $N^{x}$ play the role of Lagrange multipliers in the Hamiltonian formalism. Here $\ct$ is conformal time, and the scale factor is $a^{2}=g_{xx}=e^{2\phican}$. For the study of cosmological perturbations, the theory \eqref{Lcovpre} is coupled to a scalar inflaton field $\inflaton$ with action
\be\label{SX}
\SS_{\infsub}=\frac{1}{2}\int \sqrt{-g}\,\left[  -(\nabla\inflaton)^2 - \VV(\inflaton) \right]
\ee
The potential $\VV(\inflaton)$ contributes to the gravitational equations of motion by combining with the cosmological constant as $\Lambda+(\gamma^{2}/4)\VV(\inflaton)$. The non-zero conjugate momenta of the combined gravity and matter system (with total action \eqref{Lcovpre} plus \eqref{SX}) are
\begin{equation}\label{momenta}
\begin{aligned}
\pi_{\Laux}&=\frac{4}{\gamma^{2}N^{\ct}}\Bigl( (\Laux+\phican)'-N^{x}\px(\Laux+\phican) - \px N^{x} \Bigr)\\
\pi_{\phican}&=\frac{4}{\gamma^{2}N^{\ct}} \Bigl( \Laux'-N^{x}\px\Laux \Bigr)\\
\pi_{\inflaton}&=\frac{1}{N^{\ct}} \Bigl( \inflaton'-N^{x}\px\inflaton \Bigr)
\end{aligned}
\end{equation}
where a prime denotes the conformal time derivative $\partial_{\ct}$. The Hamiltonian and momentum constraints are then written as
\begin{equation}\label{constraints}
\begin{aligned}
\HH &= \frac{\gamma^2}{4}\Bigl( -\half\pi_\phican^2 + \pi_\phican\pi_\Laux \Bigr) + \frac{2}{\gamma^2} \Bigl( (\partial_x\Laux)^2 + 2 (\partial_x\phican)(\partial_x\Laux) - 2\partial_x^2\Laux + \Lambda e^{2\phican} \Bigr)\\
&\qquad{\;}\qquad{\;}\qquad{\;}\qquad{\;}\qquad{\;}\qquad{\;}\qquad +\half\pi_{\inflaton}^{2}+\half(\px\inflaton)^{2}+\half e^{2\phican}\VV(\inflaton)\\
\mathcal{P}&=\pi_{\Laux}\px\Laux+\pi_{\phican}\px\phican-\px\pi_{\phican}+\pi_{\inflaton}\px\inflaton
\end{aligned}
\end{equation}
The vanishing of these constraints will be imposed on cosmological perturbations to give the constrained effective action at each order in perturbation theory.

\subsubsection*{Global shift symmetry}
The existence of the topological Euler characteristic \eqref{Euler} gives rise to an important symmetry of the extended Liouville action \eqref{Lcovpre}. The coupling of the scalar curvature to the auxiliary field $\Laux$ is non-trivial only when $\Laux$ is not a constant. In other words, shifting $\Laux$ by a constant changes the action \eqref{Lcovpre} only by a multiple of the Euler characteristic, which does not affect the equations of motion. We will see in section \ref{consistencysec} that in cosmological gauges such as spatially flat gauge or constant inflaton gauge, the shift symmetry of $\Laux$ becomes an invariance under global rescalings (Weyl transformations) of the metric. This symmetry will turn out to be crucial in deriving Maldacena's consistency relation for the three-point function of curvature perturbations in 2d.

\subsection{Interpretation in worldsheet string theory}
Before moving on, let us briefly mention the connection between extended Liouville theory and more familiar theories of gravity in 2d. Firstly, the nonlocal Polyakov action for 2d gravity induced by minimally coupled matter \cite{Polyakov:1987zb} is recovered from \eqref{Lcovpre} upon eliminating $\Laux$ by its equation of motion \eqref{chieom}. Secondly, the extended Liouville theory \eqref{Lcovpre} reduces in conformal gauge to regular timelike Liouville theory. To fix conformal gauge, the metric is written as $g_{ab}=e^{2\phican}\ghat_{ab}$, but without assuming $\ghat_{xx}=1$ as in \eqref{metparam}. The Liouville field $\phican$ is allowed to fluctuate, while $\ghat_{ab}$ is fixed by gauging diffeomorphisms. The scalar curvature decomposes into
\begin{equation}\label{R2Rhat}
\sqrt{-g}\,R = \sqrt{-\ghat}\, \bigl( \hat R - 2\hat\nabla^{2}\phican \bigr)
\end{equation}
where all hatted quantities are constructed from $\ghat$. The action \eqref{Lcovpre} becomes
\begin{equation}\label{Sgravhat}
\SS_\grav  = \frac{2}{\gamma^2} \int \sqrt{-\hat g}\left[ (\hat\nabla\phican)^2 -(\hat\nabla\Theta)^2 + Q\hat\R (\phican -\Theta)   - {\Lambda}e^{2\phican} \right]  
\end{equation}
where $\Theta\equiv\Laux+\phican$, and $Q=1$ at the classical level. The vanishing trace component of the stress tensor, equation \eqref{phieom}, is now the $\phican$ equation of motion; hence \eqref{Sgravhat} with fixed reference metric $\ghat$ is classically conformally invariant. Imposing the $\Laux$ equation of motion \eqref{chieom}, and using \eqref{R2Rhat}, fixes $\Theta=0$ up to nonlocal terms in $\ghat$. The remaining terms in \eqref{Sgravhat} are precisely the action for timelike Liouville theory \cite{Martinec:2014uva} --- hence they are classically equivalent in conformal gauge.

The action \eqref{Sgravhat} may also be coupled to additional matter fields to give a nonlinear sigma model with vanishing conformal anomaly \cite{Martinec:2014uva}. The Liouville field $\phican$ becomes the time coordinate in target space, and the coupling to $\hat R$ represents a null linear dilaton. This causes a cancellation between the conformal improvements of $\phican$ and $\Theta$, so the gravity sector has central charge $c_{\phican}+c_{\Theta}=2$. Twenty-four additional scalar fields are required to cancel the conformal anomaly of the Faddeev-Popov ghosts. The resulting nonlinear sigma model describes bosonic string propagation in the critical dimension $D=26$, in the presence of a null linear dilaton, and a tachyon condensate generating the worldsheet cosmological constant $\Lambda$ (and any potential for the matter fields, such as a slow roll inflaton potential).

The above discussion of the conformal anomaly takes place in conformal gauge. In this paper, our interest will instead be the classical theory in cosmological gauges, where $\ghat_{ab}$ fluctuates. Matter fields other than the inflaton \eqref{SX} are surplus to our requirements, as the conformal anomaly does not arise in the context of the classical perturbation theory developed here. In any case, additional minimally coupled fields would be spectators in the analysis of cosmological perturbations.

\section{Slow roll expansion and quadratic fluctuation action}\label{quadsec}
We now explore the basic properties of cosmological perturbations in extended Liouville theory, emphasizing the strong parallel with the scalar sector in 4d Einstein gravity. This is partly a review of material presented in \cite{Martinec:2014uva}, and sets the stage for the analysis of non-Gaussianity in the following sections.

\subsection{Slow roll background}
To get started, we must describe the background spacetime on which cosmological perturbations propagate. We are thus interested in spatially homogeneous solutions of the equations of motion of the combined gravity and matter system. The background metric is taken to be the \dS metric
\begin{equation}\label{dSmetric}
ds^{2} = e^{2\phib}\( -d\ct^{2}+dx^{2} \)
\end{equation}
where $-\infty<\ct<0$ is conformal time. A tilde denotes a spatially homogeneous background field, such as $\phib=\phib(\ct)$. The scale factor is identified as $a(\ct)=e^{\phib(\ct)}$. Consider first the equation of motion for $\Lauxb(\ct)$, which is the reduction of \eqref{chieom} in the \dS metric \eqref{dSmetric}:
\begin{equation}
0 = \Lauxb'' + \phib''
\end{equation}
Equation \eqref{R2Rhat} was used to find the contribution of $\phib$ to the curvature. Recall that a prime denotes the conformal time derivative $\partial_{\ct}$. We choose to consider only backgrounds $\Lauxb(\ct)$ satisfying
\begin{equation}\label{chisol}
0 = \Lauxb' + \phib'
\end{equation}
This choice helps facilitate the comparison of cosmological perturbations in 1+1d with those in 3+1d, as will be seen in section \ref{pertsec}. Eliminating $\Lauxb'(\ct)$ via \eqref{chisol}, the remaining background equations of motion and Hamiltonian constraint become
\begin{equation}\label{cteoms}
\begin{aligned}
0 &= -\phib'' +  e^{2\phib}\bigl[ \Lambda + (\gamma^{2}/4) \VV(\infb) \bigr] \\
0 &= \infb'' + \half e^{2\phib}\VV_{,\infsub} (\infb) \\
0 &= -(\phib')^{2} + (\gamma^{2}/4)(\infb')^{2} + e^{2\phib}\bigl[ \Lambda + (\gamma^{2}/4) \VV(\infb) \bigr]
\end{aligned}
\end{equation}
These are the Friedmann equations of 2d cosmology \cite{DaCunha:2003fm}. The first line is the reduction of the $\phican$ equation of motion \eqref{phieom}. The second line is the matter equation of motion arising from the action \eqref{SX}, and the third line is the Hamiltonian constraint \eqref{constraints}. The momentum constraint is satisfied trivially by spatially homogeneous backgrounds. Of course, only two of the Friedmann equations \eqref{cteoms} are unique, a situation familiar from 4d. In the absence of matter, the \dS metric \eqref{dSmetric} solves the equations with $\phib(\ct)$ given by
\begin{equation}
\sqrt{\Lambda}\, e^{\phib} = -1/\ct
\end{equation}

The slow roll expansion is most easily described in coordinate time $\pht$, defined by
\begin{equation}
d\pht=a(\ct)\, d\ct~,\qquad -\infty<t<\infty
\end{equation}
The Friedmann equations \eqref{cteoms} are expressed in coordinate time as
\begin{equation}\label{phteoms}
\begin{aligned}
0 &= -\ddot{\phib}-\dot{\phib}^{2} + \Lambda  + (\gamma^{2}/4) \VV(\infb) \\
0 &= \ddot{\infb} + \dot{\phib}\dot{\infb} + \half  \VV_{,\infsub}(\infb) \\
0 &= -\dot{\phib}^{2} + \Lambda  + (\gamma^{2}/4) \bigl[ \dot{\infb}^2 
 + \VV(\infb)\bigr]
\end{aligned}
\end{equation}
where a dot denotes $\partial_{\pht}$. The Hubble parameter is simply $H=\dot a/a=\partial_{t}\phib$. The slow roll expansion may now be developed following \cite{Baumann:2009ds}. The dimensionless slow roll parameters are defined as
\begin{equation}\label{SRparams}
\epsilon \equiv \frac{d}{d\pht}\(\frac{1}{H}\)~,\quad
\eta \equiv \frac{\dot{\epsilon}}{H\epsilon}~,\quad
\kappa \equiv \frac{\dot{\eta}}{H\eta}~,\quad
\delta \equiv -\frac{\ddot{\infb}}{H\dot{\infb}}
\end{equation}
Combining the Friedmann equations \eqref{phteoms} to eliminate $\VV(\inflaton)$ gives
\begin{equation}\label{SRvel}
\epsilon  = \frac{\gamma^{2}}{4} (\dot{\infb}/\dot{\phib})^{2} ~,\qquad
\eta = 2(\epsilon-\delta) 
\end{equation}
In the slow roll approximation, both $\epsilon$ and $|\delta|$ are assumed to be much less than one --- typically of order a few percent. Then $|\eta|$ is also small, signifying a long-lived phase of slow roll inflation. The slow roll parameters may also be expressed in terms of the matter potential as
\begin{equation}\label{SRpot}
\epsilon = \frac{1}{\gamma^{2}} \( \frac{\VV_{,\infsub}}{\VV} \)^{2} \(\frac{1-\epsilon}{1-\delta}\)^{2}~,\qquad
\eta = \frac{4}{\gamma^{2}} \biggl[ \frac{\VV_{,\infsub\infsub}}{\VV}-\(\frac{\VV_{,\infsub}}{\VV}\)^{2}\, \biggr] + \cdots
\end{equation}
This is all precisely analogous to the situation in 4d.

In what follows, explicit slow roll trajectories for the background fields will not be required. Nevertheless, some simple examples are now given for completeness. One exact solution consists of a quadratic potential of frequency $\omega$, and an inflaton trajectory which is overdamped:
\begin{equation}
\begin{aligned}
\VV(\inflaton) &= \half\omega^{2}\inflaton^{2} + \frac{2\omega}{\gamma}\sqrt{2\Lambda+\omega^{2}}\,\inflaton \\
\infb(\pht) &= \frac{\sqrt{2}}{\gamma}\,\omega \pht+\Xb_{0} \\
\phib(\pht) &= -\frac{1}{4}\omega^{2}\pht^{2} - b\pht + \phib_{0} \\
b &= \sqrt{\Lambda+\omega^{2}/2}+\frac{\gamma}{\sqrt{8}}\omega\infb_{0}
\end{aligned}
\end{equation}
Recall that coordinate time ranges over $-\infty<t<\infty$. The scale factor $e^{\phib}$ expands prior to the moment $t_{c}=-2b/\omega^{2}$, after which it contracts. The effective cosmological constant at the minimum of the potential may always be taken positive by adjusting $\Lambda$. There is an infinite era (in coordinate time $t$) of slow roll inflation when $t<(t_{c}-\sqrt{2}/\omega)$. This era may be cut out and smoothly extended into \dS space.

Another example is the exponential potential $\VV(\inflaton)=e^{-\beta\inflaton}$. In this case the explicit trajectories for $\infb$ and $\phib$ are most easily found from the conformal time equations \eqref{cteoms}, and the slow roll conditions are satisfied for $|\beta|/\gamma\ll1$. One may also construct approximate solutions for a quartic potential, following exactly those in \cite{Linde:1983gd}.

\subsection{Perturbations and gauge transformations}\label{pertsec}
Cosmological perturbations are now introduced, with discussion of their gauge transformations and effective action. The theory of gauge-invariant perturbations was introduced in \cite{Bardeen:1980kt}, and reviewed in \cite{Mukhanov:1990me,Baumann:2009ds}. In 3+1 dimensions, perturbations of the metric are grouped into scalar, vector, and tensor representations of the Lorentz group. Our interest is in the scalar sector, which is precisely related to the perturbative degrees of freedom of extended Liouville theory. When perturbing the metric about a 4d \dS background, the scalar degrees of freedom are commonly parametrized as \cite{Baumann:2009ds}
\be\label{ds4d}
ds^2_{4d} = a^2 \Bigl\{ -(1+2\Phi)d\ct^2 + 2B_{,i} dx^i d\ct + \bigl[(1-2\Psi)\delta_{ij}+2E_{,ij}\bigr]dx^i dx^j \Bigr\}
\ee
There are four scalar degrees of freedom in the metric ($\Phi,\Psi,B$, and $E$), which parametrize the trace component, and curl-free contributions to the shift vector and space-space metric. The vector and tensor sectors --- which are not considered here --- consist of divergence-free vector and transverse traceless tensor contributions, which cannot be expressed as derivatives of scalar functions. The behavior of \eqref{ds4d} under diffeomorphisms is of interest for constructing gauge-invariant quantities. The scalar sector is closed under diffeomorphisms of the form
\be\label{coordtransf4d}
\ct\to \ct+\xi^\ct(\ct,\bx)  ~~, \quad x^i\to x^i +\partial^i\xi(\ct,\bx)
\ee
where $\xi^{\ct}$ and $\xi$ are scalar functions. For each quantity $\Sigma$ which is split into background plus perturbation as $\Sigma=\tilde\Sigma+\bm\sigma$, we are interested in the linearized gauge transformation
\begin{equation}\label{Lie}
\bm\sigma \to \bm\sigma + \mathsterling_{\xi}\bigl( \tilde\Sigma+\bm\sigma \bigr)
\end{equation}
In the case of the scalar sector, $\bm\sigma$ ranges over $\Phi$, $\Psi$, $B$, and $E$. The entire gauge transformation is assigned to the perturbation, leaving the background invariant. This is appropriate for describing the behavior of perturbations about a fixed background, and coincides with the gauge redundancy of the path integral. In this language, a time-dependent background breaks the gauge symmetry under time reparametrizations. The part of \eqref{Lie} which is homogeneous in $\bm\sigma$ --- that is, the first order Taylor expansion of $\bm\sigma(x+\xi)$ --- is trivial and will be omitted for brevity. The gauge transformation of the scalar perturbations in \eqref{ds4d} is then \cite{Mukhanov:1990me,Baumann:2009ds}
\be\label{fieldtransf4d}
\begin{aligned}
\Phi &\to \Phi + \partial_\ct\xi^\ct + \frac{a'}{a} \xi^\ct\\
\Psi &\to \Psi - \frac{a'}{a} \xi^\ct  \\
B &\to B - \xi^\ct + \xi'\\
E &\to E + \xi
\end{aligned}
\ee

Turning now to extended Liouville theory, we again split all fields into spatially homogeneous backgrounds (denoted by a tilde) plus inhomogeneous perturbations (bold lower case). Using the metric parametrization \eqref{metparam} and adopting the \dS background \eqref{dSmetric}, the perturbative degrees of freedom in the gravitational sector are
\be\label{fieldperts}
\begin{aligned}
\phican  &= \phib(\ct) + \phipert(\ct,x) \\
N^\ct  &= 1 + \ntpert(\ct,x) \\
N^x  &= \nxpert(\ct,x) \\
\Laux &= \Lauxb(\ct) + \Lauxpert(\ct,x) 
\end{aligned}
\ee
There are only three metric degrees of freedom in 2d, but the auxiliary field perturbation $\Lauxpert$ provides a fourth. For explicit comparison with \eqref{ds4d}, the perturbed line element is
\begin{equation}\label{ds2d}
ds^{2}_{2d} = e^{2\phib}\Bigl\{ -(1+2\phipert+2\ntpert)d\ct^{2} + 2\nxpert dxd\ct + (1+2\phipert)dx^{2} \Bigr\}
\end{equation}
These are all scalar perturbations; the vector and tensor sectors are absent in 2d. Comparison of coefficients in \eqref{ds2d} and \eqref{ds4d} suggests an obvious identification between certain linear combinations of metric perturbations. For instance, $\Phi$ should be identified with $\phipert + \ntpert$, and this linear combination merely reflects the differing metric parametrizations adopted in different dimensions. The non-trivial part of the identification concerns the auxiliary field perturbation $\Lauxpert$, which acts as the fourth gravitational degree of freedom in extended Liouville theory. To understand this, it helps to consider the 2d analog of the gauge transformations \eqref{fieldtransf4d}. In 2d, all infinitesimal diffeomorphisms are locally of the form \eqref{coordtransf4d}, and the linearized gauge transformation of the perturbations is
\begin{equation}\label{fieldtransf}
\begin{aligned}
\phipert &\to \phipert + \phib'\xi^\ct + \px^{2}\xi\\
\ntpert &\to \ntpert + \partial_\ct{\xi^\ct}  - \px^{2} \xi\\
\nxpert &\to \nxpert - \partial_x\xi^\ct + \px\xi'\\
\Lauxpert &\to \Lauxpert - \phib' \xi^\ct
\end{aligned}
\end{equation}
For comparison with \eqref{fieldtransf4d}, recall that $a=e^{\phib}$, so $\phib'=a'/a$. In the last line of \eqref{fieldtransf}, the background condition $\Lauxb'+\phib'=0$ \eqref{chisol} allowed us to write the gauge transformation of the auxiliary scalar field ($\Lauxpert\to\Lauxpert+\Lauxb'\xi^{\ct}$) in terms of $\phib'$. In such backgrounds, then, the gauge transformation of $\Lauxpert$ is identical to that of $\Psi$ in 4d, allowing us to identify these two perturbations. The complete map between extended Liouville perturbations and scalar perturbations in 3+1 Einstein gravity is then
\be\label{pertmap}
\begin{aligned}
\phipert &~ \leftrightarrow~ -\Psi + \vec\partial^{\,2} E\\
\ntpert &~\leftrightarrow~ \Phi + \Psi - \vec\partial^{\,2} E \\
\int^{x}dx'\, \nxpert &~ \leftrightarrow~ B\\
\Lauxpert &~ \leftrightarrow~ \Psi
\end{aligned}
\ee
Fluctuations of the scalar inflaton in 2d and 4d are denoted by
\begin{equation}
\inflaton  = \infb(\ct) + \infpert(\ct,x) \nonumber
\end{equation}
Their gauge transformation is
\begin{equation}\label{infgaugetrans}
\infpert \to \infpert + \infb'\xi^{\ct}
\end{equation}
and $\infpert_{2d}$ is identified with $\infpert_{4d}$.

Gauge-invariant combinations of metric and matter perturbations are constructed in parallel with well-known 4d results. The introduction of the scalar inflaton makes for five scalar degrees of freedom in both dimensions, which may be arranged into three gauge-invariant linear combinations. The Bardeen potentials in 4d are \cite{Bardeen:1980kt}
\begin{equation}
\begin{aligned}
\Psi_{\rm B} &= \Psi + \frac{a'}{a}(E'-B) \\
\Phi_{\rm B} &= \Phi - \frac{1}{a}\, \partial_{\ct}\bigl[ a(E'-B) \bigr]
\end{aligned}
\end{equation}
and their 2d counterparts follow from the map \eqref{pertmap}. In order to isolate $E$, one must spatially integrate twice the first two lines of \eqref{pertmap}, reflecting the different parametrization of the space-space metric. The third gauge-invariant variable is a linear combination of $\Psi$ and the inflaton perturbation $\infpert$, and is known as the Mukhanov-Sasaki variable \cite{Mukhanov:1988jd,Sasaki:1986hm}. For convenience in writing the quadratic effective action, this variable contains a dimension-dependent power of the scale factor, which has no effect on its gauge transformation. To facilitate a unified presentation in the following discussion, the spacetime dimension will be denoted by $d$. Only the values $d=2$ and $d=4$ are considered:
\begin{equation}
d = 2n \qquad (n=1,2)
\end{equation}
The gauge-invariant Mukhanov-Sasaki variable is then
\begin{equation}\label{vdef}
v = a^{n-1}\infpert + z\Lauxpert
\end{equation}
subject to the identification $\Lauxpert\leftrightarrow\Psi$. The quantity $z$ is given by
\begin{equation}\label{zdef}
z(\ct) = a^{n-1}(\dot\infb/H) = (2/\gamma)a^{n-1}\sqrt{\epsilon}
\end{equation}
To interpret the second equality in 4d, recall that $2/\gamma$ plays the role of the Planck mass $m_{p}$.

\subsubsection*{Cosmological gauge choices}
The identifications \eqref{pertmap} mean that any gauge choice for the scalar sector in 4d has an analog in extended Liouville theory. Two popular gauges which will be of particular interest in this paper are \cite{Baumann:2009ds}
\begin{align}
\qquad \Psi = 0 = E~&\lr~ \Lauxpert = 0 = \phipert &\text{(spatially flat gauge)}\label{sfgauge}\\
\qquad \infpert = 0 = E~&\lr~ \infpert = 0 = \Lauxpert + \phipert  &\text{(constant inflaton gauge)}\label{cfgauge}
\end{align}
Both of these gauge choices completely fix local diffeomorphisms. In spatially flat gauge in 3+1d, the local coordinates are chosen to set to zero the curvature of spatial slices. Spatial slices in 1+1d don't have curvature, but still the spatial metric is fixed to the identity in spatially flat gauge. Constant inflaton gauge has the same interpretation in both dimensions --- the time-dependent inflaton background is used to set our clock.

\subsection{Quadratic fluctuation action}\label{Mukhanovsec}
The structure of the quadratic fluctuation action for extended Liouville theory plus the scalar inflaton follows precisely the familiar Mukhanov-Sasaki action in 3+1 dimensions \cite{Mukhanov:1988jd,Sasaki:1986hm}. At each order in perturbation theory (quadratic, cubic, etc.), a classical effective action is constructed by adjusting the lapse and shift fluctuations to enforce the constraints.\footnote{This is not the only possible way to proceed. One may also construct perturbation theory in the Hamiltonian formalism, where the lapse and shift are Lagrange multipliers whose value is arbitrary. The constraints are then solved by adjusting dynamical variables and/or their momenta. Translating back to the Lagrangian formalism, one finds at quadratic order the lapse and shift appear only as total derivatives. They drop out of the action while remaining arbitrary, leaving the familiar Mukhanov-Sasaki action. In some sense this procedure is more natural than that employed in the text, because it makes clear why the result \eqref{Meffact} depends only on $v$, and not $\Phi_{\rm B}$ or $\Psi_{\rm B}$ --- the latter contain arbitrary Lagrange multipliers.} The constrained lapse and shift fluctuations $\ntpert$ and $\nxpert$ are functions of the dynamical variables $\Omega$, whose fluctuations are collectively denoted $\bm\omega$. This results in an expansion
\begin{equation}\label{naexp}
\bm n^{a}(\bm\omega) = \bm n^{a}_{1}(\bm\omega) + \bm n^{a}_{2}(\bm\omega) + \cdots \qquad (a=\ct,x)
\end{equation}
where the subscript denotes the order in $\bm\omega$. The quadratic action is composed of terms of the form $\bm\omega^{2}$, $\bm\omega\, \bm n_{1}(\bm\omega)$, or $\bm n_{2}(\bm\omega)$. The coefficient of the latter is the vanishing background constraints, so we only need to know the first order lapse and shift $\bm n^{a}_{1}(\bm\omega)$.

In the case of extended Liouville theory plus the scalar inflaton, the Hamiltonian and momentum constraints were given in \eqref{constraints}. Their expansion to first order in fluctuations is
\begin{equation}\label{constraintpert}
\begin{aligned}
(\gamma^{2}/4)\,\cH_{1} &= \phib'^{\,2}(1-\epsilon) \bigl(\bm n^{\ct}_{1}+\phipert\bigr) + \phib'\bigl(\px\bm n^{x}_{1} - \phipert'\bigr) + \px^{2}\Lauxpert + \frac{\gamma^{2}}{4}\Bigl( \infb'\infpert' + \half e^{2\phib}\VV_{,\infsub}(\infb)\infpert \Bigr) \\
(\gamma^{2}/4)\,\cP_{1} &= - \phib' \px\( \bm n^{\ct}_{1} + \phipert + \Lauxpert'/\phib' - \frac{\gamma}{2}\sqrt\epsilon\, \infpert \)
\end{aligned}
\end{equation}
The equations $\cH_{1}=0=\cP_{1}$ are solved to give
\be\label{ntnx}
\begin{aligned}
\bm n^{\ct}_{1} &= -\phipert - \Lauxpert'/\phib' + \frac{\gamma}{2}\sqrt\epsilon\, \infpert \\
\bm n^{x}_{1} &= \px\Lauxpert/\phib' + \psi 
\end{aligned}
\ee
where
\begin{equation}\label{psix}
\px\psi = \phipert' + (1-\epsilon)\Lauxpert' - \frac{\gamma}{2}\epsilon\bigl(\infpert/\sqrt\epsilon\bigr)'
\end{equation}
The fact that $\psi$ is the subject of a differential equation has the effect of introducing spatial nonlocality into the constrained effective action. The quadratic action, however, may be expressed in terms of $\px\psi$ only, using integration by parts; at cubic order we will see the nonlocality of $\psi$ is genuine. It is also gauge-invariant, because $\px\psi$ changes only by a total $x$-derivative under gauge transformations.

To arrive at the constrained quadratic action, one expands the extended Liouville action \eqref{Lcovpre} plus scalar inflaton \eqref{SX} to second order in fluctuations, substitutes \eqref{ntnx}, and integrates by parts. The final result is the gauge-invariant Mukhanov-Sasaki action \cite{Mukhanov:1988jd,Sasaki:1986hm}
\bea
\label{Meffact}
\SS_{2} = \frac{1}{2}\int\biggl( v'^{\,2}-(\partial_x v)^2 + \frac{z''}{z} v^2 \biggr)
\eea
The form of this action is identical in 2d and 4d. The Mukhanov-Sasaki variable $v$ is effectively a free scalar field with time-dependent mass $m^{2}=z''/z$. In constant inflaton gauge \eqref{cfgauge}, where $v$ given by \eqref{vdef} reduces to $-z\phipert\sim-\epsilon^{1/2}\phipert$, the action \eqref{Meffact} is manifestly suppressed by $\epsilon$. For a more detailed derivation of \eqref{Meffact} in extended Liouville theory, see \cite{Martinec:2014uva}.

\subsection{Mode functions}\label{modesec}
We now investigate the equation of motion for the Mukhanov-Sasaki action, solving for the spatial momentum mode functions of $v(\ct,\bx)$. We continue to denote the spacetime dimension by $d=2n$. The equation of motion following from the action \eqref{Meffact} is
\begin{equation}\label{veom}
\Bigl( \partial^{2} + \frac{z''}{z} \Bigr) v = 0
\end{equation}
When the field $v(\ct,\bx)$ is expanded into spatial momentum modes as
\begin{equation}\label{vexp}
v(\ct,\bx) = \int\frac{d^{d-1}k}{(2\pi)^{d-1}} \Big( a(\bk)v_{k}(\ct)e^{i\bk\cdot\bx} + \text{h.c.} \Big)
\end{equation}
equation \eqref{veom} becomes the Mukhanov-Sasaki equation for the modes $v_{k}(\ct)$:
\begin{equation}\label{Mukhanoveq}
v_{k}'' + \Bigl( k^{2}-\frac{z''}{z} \Bigr) v_{k} = 0 
\end{equation}
Following \cite{Baumann:2009ds}, we proceed to solve this equation by expanding the mass term $z''/z$ in slow roll. Differentiating $z(\ct)$ given by \eqref{zdef} and comparing with the slow roll parameters \eqref{SRparams}, one finds the exact expressions
\begin{equation}\label{SRz1}
\begin{aligned}
\frac{z'}{z} & = aH\Bigl[ n-1 + \half\eta \Bigr] \\
\frac{z''}{z} & =(aH)^{2}\Bigl[ (n-1)(n + \eta-\epsilon) + \half\eta\Bigl( 1 + \half\eta - \epsilon + \kappa \Bigr) \Bigr]
\end{aligned}
\end{equation}
Note that the mass term $z''/z$ vanishes in 2d at zeroth order in slow roll. This means that in 2d \dS space, the Mukhanov-Sasaki variable is a massless scalar field. The prefactor $aH=\phib'$ in \eqref{SRz1} is expressed in terms of the slow roll parameter $\epsilon$ as
\begin{equation}
\frac{1}{aH} = -\ct(1-\epsilon)
\end{equation}
which satisfies $aH\to\infty$ as $\ct\to0^{-}$. Substituting in \eqref{SRz1} gives the exact expressions
\begin{equation}\label{SRz2}
\begin{aligned}
\frac{z'}{z} & = -\frac{1}{\ct(1-\epsilon)}\Bigl[ n-1 + \half\eta \Bigr] \\
\frac{z''}{z} & =\frac{1}{\ct^{2}(1-\epsilon)^{2}} \Bigl[ (n-1)(n + \eta-\epsilon) + \half\eta\Bigl( 1 + \half\eta - \epsilon + \kappa \Bigr) \Bigr]
\end{aligned}
\end{equation}
Although $z=0$ in \dS space (where all slow roll parameters vanish), the ratios \eqref{SRz2} are non-vanishing in de Sitter, except in 2d. The Mukhanov-Sasaki equation \eqref{Mukhanoveq} may now be written in the form of a Bessel equation:
\begin{equation}\label{Beq}
 v_{k}'' + \( k^{2}-\frac{\nu^{2}-1/4}{\ct^{2}} \) v_{k}  = 0
\end{equation}
The order $\nu$, which is constant only to first order in slow roll, is given by
\begin{equation}
\nu = n - \half + (n-1)\epsilon + \half\eta + \cdots
\end{equation}
The ellipsis represents terms of higher order in slow roll; by ignoring time derivatives of $\nu$ we are already ignoring such terms. The first order solutions to \eqref{Beq} are
\begin{equation}\label{vksol}
v_{k}(\ct) = \sqrt{-\ct}\, H_{\nu}^{(1,2)}(-k\ct)
\end{equation}
The Bunch-Davies condition states that in the far past, when the mode $\bk$ is much smaller than the horizon, the vacuum should be the usual Minkowski vacuum. In other words, if we use the vacuum annihilated by $a(\bk)$ in the mode expansion \eqref{vexp}, then the modes $v_{k}$ must have positive energy according to the Minkowski space representation $E=i\pt$. The asymptotic behavior of the Hankel function at early times ($-k\ct\to\infty$) follows from
\begin{equation}
H_{\nu}^{(1,2)}(x) \sim \sqrt{\frac{2}{\pi x}} \,\exp \pm i\Bigl( x-\half\nu\pi -\frac{1}{4}\pi \Bigr)\qquad (x\to\infty)
\end{equation}
Hence the normalized Bunch-Davies solution is given by the Hankel function of the first kind:
\begin{equation}
v_{k}^{(+)}(\ct) = \half e^{+i\pi(\nu/2+1/4)}\sqrt{-\pi \ct}\,H_{\nu}^{(1)}(-\mk\ct)  
\end{equation}
The overall phase is arbitrary and has been chosen for convenience. The Bunch-Davies vacuum will be adopted exclusively from here, and the superscript $(+)$ will be dropped.

In constant inflaton gauge \eqref{cfgauge}, the Mukhanov-Sasaki variable \eqref{vdef} reduces to $v=-z\phipert \leftrightarrow z\Psi$. The comoving curvature perturbation\footnote{The comoving curvature perturbation in 3+1d describes the curvature of spatial slices in constant inflaton gauge. In 1+1d, spatial slices don't have curvature, so the name ``curvature perturbation'' is something of a misnomer. Perhaps $\zeta$ is best thought of in 1+1d as a perturbation of the scale factor. For the same reason, spatially flat gauge should perhaps be called ``constant scale factor gauge'' in 1+1d.} $\zeta$ is defined by $v=-z\zeta$, or
\begin{equation}
\zeta_{2d}=\phipert~,\qquad \zeta_{4d}=-\Psi
\end{equation}
There are differing sign conventions for $\zeta$ in the literature; our choice corresponds to a perturbed scale factor $e^{\phib+\zeta}$  in constant inflaton gauge. The explicit curvature mode functions at lowest order in slow roll are
\begin{align}
\zeta_{k}(\ct) &= - \frac{1}{z(\ct)} \frac{e^{-ik\ct}}{\sqrt{2k}} + \cdots &(d=2) \label{zetak2d} \\
\zeta_{k}(\ct) &=  \frac{-i}{z(\ct)(-\ct)} \frac{e^{-ik\ct}}{\sqrt{2k^{3}}} \(1+ik\ct\) + \cdots  &(d=4) \label{zetak4d}
\end{align}
Note that the denominators carry only a weak time dependence, with an exponent proportional to the slow roll parameters. This is because $z(\ct)$ is given to first order in slow roll by integrating the first order expansion of \eqref{SRz2}, pretending the slow roll parameters are constant:
\begin{equation}\label{zct0}
z(\ct) =z(\ct_{0}) (\ct/\ct_{0})^{\half-\nu}+\cdots
\end{equation}
Here $\ct_{0}$ could be any reference time, so the product $z(\ct)(-\ct)^{\nu-\half}=z(\ct)(-\ct)^{n-1+\cdots}$ is constant to first order in slow roll. Note also that the modes \eqref{zetak2d} and \eqref{zetak4d} freeze out in physical time $t$ with a dimension-dependent power of $k/aH=-k\ct$:
\begin{equation}\label{freezeout}
\pt\zeta_{k} \propto (-k\ct)^{n} \zeta_{k} \qquad (-k\ct\ll1)
\end{equation}
 In particular, $\partial_{\ct}\zeta_{k}=a\,\pt\zeta_{k}=(-1/H\ct)\pt\zeta_{k}$ does not vanish in 2d as $\ct\to0^{-}$, so one must be careful to assume only the vanishing of $\pt\zeta_{k}$ on superhorizon scales. This slower rate of freeze out may be traced back to the vanishing in 2d of the effective mass $m^{2}=-z''/z$ in the Mukhanov-Sasaki equation \eqref{Mukhanoveq}.

\subsection{Power spectrum}\label{powersec}
The power spectrum of scalar curvature perturbations, $P_{\zeta}(k)$, is the Fourier transform of the two-point function:
\begin{equation}
\langle0| \zeta(\ct,\bx)\zeta(\ct,\bx') |0\rangle = \int\frac{d^{d-1}k}{(2\pi)^{d-1}} e^{i\bk\cdot(\bx-\bx')} P_{\zeta}(k)
\end{equation}
Expanding the left side into mode functions, one finds
\begin{equation}
P_{\zeta}(k) = \big| \zeta_{k}(\ct) \big|^{2}
=\frac{\pi}{4} \(\frac{-\ct}{z^{2}}\) \big|H_{\nu}^{(1)}(-\mk\ct)\big|^{2}
\end{equation}
To evaluate this on superhorizon scales ($-k\ct\ll1$), we need the late-time behavior of the Hankel function:
\begin{equation}
H_{\nu}^{(1)}(x) \sim -(i/\pi)\Gamma(\nu)(x/2)^{-\nu} \qquad (x\to0^{+})
\end{equation}
The power spectrum on superhorizon scales is then
\begin{equation}\label{P1}
P_{\zeta}(k) = 2^{2\nu-2} \Bigl[z(\ct)(-\ct)^{\nu-\half}\Bigr]^{-2} \frac{1}{k^{2\nu}}
\qquad (-k\ct\ll1)
\end{equation}
The factor in brackets is constant by \eqref{zct0}, so $1/k^{2\nu}=1/k^{d-1+\cdots}$ is the complete momentum dependence of the power spectrum at first order in slow roll. The spectral tilt (the deviation of the exponent from the scale-invariant value $(d-1)$\,) is
\begin{equation}\label{tilt}
(n_{s}-1)  = (d-1) - 2\nu =  -2(n-1)\epsilon - \eta + \cdots
\end{equation}
Another useful expression for the power spectrum is obtained by specializing the bracket in \eqref{P1} to the moment $\ct_{*}(k)$ when the mode $k$ crosses the horizon, given by
\begin{equation}
\ct_{*}(k) = -1/k
\end{equation}
Evaluating the constant bracket in \eqref{P1} at this time (using the definition \eqref{zdef} of $z(\ct)$) leads to
\begin{equation}\label{P2}
P_{\zeta}(k) = 2^{d-5}\gamma^{2}\, \frac{H_{*}^{d-2}}{\epsilon_{*}}\, \frac{1}{k^{d-1}}
\end{equation}
where $H_{*}$ and $\epsilon_{*}$ are evaluated at horizon crossing. The spectral tilt is now contained in the momentum dependence of $H_{*}$ and $\epsilon_{*}$. The form \eqref{P2} of the power spectrum is identical in 2d and 4d (recall that $\gamma^{2}/4$ plays the role of $1/m_{p}^{2}$). A difference is that the spectral tilt \eqref{tilt} depends only on $\eta$ in 2d; this traces back to the dimension-dependent factor of $a$ in $z(\ct)$ given by \eqref{zdef}.

This concludes the quantitative comparison of extended Liouville theory with the scalar sector of Einstein gravity at quadratic order. (As a final note, extended Liouville theory has a tensor to scalar ratio of $r=0$\,! Although, if one wished, it might be possible to mimic the 4d vector and tensor sectors by including extra degrees of freedom in 2d, with a judicious choice of action.) All equations in the remainder of this paper are specialized to $d=2$. To underline this, the degree of freedom in constant inflaton gauge will henceforth be denoted $\phipert$, although one should keep in mind that this plays the role of $\zeta$.

\subsection[Constancy of zeta outside the horizon]{Constancy of $\zeta$ outside the horizon}\label{constancysec}
Before turning to the cubic action, let us confirm the constancy of $\zeta_{2d}=\phipert$ outside the horizon, to all orders in perturbations. To do this, we begin from the full action \eqref{Lcovpre} plus \eqref{SX}, specialize to constant inflaton gauge, and neglect spatial derivatives of the dynamical fields. The key point \cite{Maldacena:2002vr} is that if the expansion of the superhorizon action in physical time derivatives of $\phipert$ begins at second order, then the equation of motion for $\phipert$ will always admit solutions with $\dot\phipert=0$. Hence it will be enough to show the zeroth and first order terms in this expansion vanish.

For ease in applying the constraints to all orders in perturbations, we begin from the Hamiltonian form of the action,
\begin{equation}\label{SHam}
\SS = \int \Bigl( \pi_{\Laux}\Laux' + \pi_{\phican}\phican' + \pi_{\inflaton}\inflaton' - N^{\ct}\cH - N^{x}\cP \Bigr)
\end{equation}
where prime continues to denote $\partial_{\ct}$. The conjugate momenta and constraints were given in \eqref{momenta} and \eqref{constraints}. In backgrounds satisfying $\Lauxb'+\phib'=0$, the constant inflaton gauge condition $\Lauxpert+\phipert=0$ means the full fields (background plus perturbation) obey $\Laux'+\phican'=0$. For simplicity we assume that $\Lambda$ has been absorbed in $\VV(\infb)$; recall that $\inflaton=\infb$ is also fixed in constant inflaton gauge. The constraints \eqref{constraints} now reduce to
\begin{equation}\label{cfconstraints}
\begin{aligned}
(N^{\ct})^{2}\, \cH &= \frac{2}{\gamma^{2}} \(-\phi'^{2} +2\phi'\px N^{x} \) + \half\infb'^{2} + \half e^{2\phican}\VV(\infb)(N^{\ct})^{2} \\
(N^{\ct})^{2}\, \cP &= -\frac{4}{\gamma^{2}} \phican' \px N^{\ct}
\end{aligned}
\end{equation}
Enforcing the vanishing of these constraints in the action \eqref{SHam} in constant inflaton gauge gives outside the horizon
\begin{equation}\label{Scfcon}
\SS_{\rm constr.} = -\int  e^{2\phican}\VV(\infb)N^{\ct} = -\int  e^{2\phib+2\phipert}\VV(\infb) \[ 1+ \ntpert(\phipert) \]
\end{equation}
Here $\ntpert(\phipert)$ is the solution of the perturbed Hamiltonian constraint. So far we have worked to all orders in field perturbations and time derivatives --- only spatial derivatives have been neglected. We now expand the Hamiltonian constraint \eqref{cfconstraints} to first order in time derivatives of $\phipert$, assuming that $\ntpert,\nxpert$ are themselves of this order. This assumption is consistent with the structure of \eqref{cfconstraints}, once the background constraints are applied. The perturbed Hamiltonian constraint is solved for $\ntpert$, giving
\begin{equation}\label{cfnt}
 \VV(\infb)\ntpert = \frac{4}{\gamma^{2}}\dot\phib \( \dot\phipert - e^{-\phib-\phipert}\px\nxpert \) 
\end{equation}
Physical time $t$ has been adopted to allow us to combine terms in the action. The measure $d\ct$ in the action is replaced by $e^{-\phib-\phipert}dt$. Substituting \eqref{cfnt} in \eqref{Scfcon} and eliminating the remaining $\VV(\infb)$ by the $\phib$ equation of motion \eqref{phteoms}, one finds
\begin{equation}
\SS_{\rm constr.}^{(<2)} = -\frac{4}{\gamma^{2}} \int e^{\phib+\phipert}\( \ddot\phib + \dot\phib^{2} + \dot\phib\dot\phipert \) 
= -\frac{4}{\gamma^{2}} \int \partial_{t}\( e^{\phib+\phipert} \dot\phib \)
\end{equation}
This does not contribute to the equation of motion, so indeed the expansion in physical time derivatives begins at second order. There are always solutions with $\dot\phipert=0$, so $\phipert$ freezes out to all orders in perturbation theory. This is consistent with our earlier finding at quadratic order in fields, equation \eqref{freezeout}. It doesn't tell us about the rate of freeze out to all orders, but this may be expected to follow the pattern at quadratic order, where freeze out was slower in 2d due to the vanishing effective mass in \dS space.

\section{Cubic fluctuation action}\label{cubicsec}

The extended Liouville action \eqref{Lcovpre} plus scalar inflaton \eqref{SX} is now expanded to cubic order in fluctuations, constrained, and massaged into a form suitable for computation of the three-point function. Our discussion follows closely that given by Maldacena \cite{Maldacena:2002vr}. We do not repeat at cubic order the manifestly gauge-invariant structure of the quadratic action \eqref{Meffact}, in terms of the Mukhanov-Sasaki variable. This would require the use of variables invariant under second-order gauge transformations, the formulation of which is highly complex (see, for example, \cite{Nakamura:2004rm}). Instead, we follow \cite{Maldacena:2002vr} by computing the cubic action first in spatially flat gauge \eqref{sfgauge}, before switching to constant inflaton gauge \eqref{cfgauge}. A comparison of these two gauges allows us to make clear in each the suppression of the cubic action by two powers of the slow roll parameter $\epsilon$. A new feature at cubic order is the spatial nonlocality of the constrained action, arising from the solution $\psi$ of the differential equation \eqref{psix}. This appears the same in each gauge, because $\px\psi$ changes only by a total $x$-derivative under gauge transformations.

The suppression of the cubic action by an additional power of $\epsilon$ over the quadratic action is the basis for perturbation theory --- the perturbative expansion is an expansion in slow roll parameters. These are assumed to be of order a few percent, hence terms higher than $O(\epsilon^{2})$ will be freely discarded in the following. 

Before beginning, we mention a helpful simplification regarding the constraints \cite{Maldacena:2002vr}. The cubic computation continues the perturbative program of ``constrain, then quantize'' begun in sections \ref{Mukhanovsec} - \ref{powersec}. Again the perturbed constraints are solved by adjusting the lapse and shift fluctuations, resulting in the order-by-order expansion \eqref{naexp}. It turns out that the cubic action, like the quadratic action, depends only on the first order lapse and shift fluctuations $\bm n^{a}_{1}(\bm\omega)$ --- the first term in \eqref{naexp}. This is because the third order term $\bm n^{a}_{3}(\bm\omega)$ is multiplied by the vanishing background constraints, while the second order term $\bm n^{a}_{2}(\bm\omega)$ appears in the cubic action as (schematically)
\begin{equation}
\(\fdd{S}{N^{2}}\Bigr|_{0}\bm n_{1}(\bm\omega)+\fdd{S}{N\delta\Omega}\Bigr|_{0}\bm\omega\)\bm n_{2}(\bm\omega)
\end{equation}
Recall that $\Omega$ runs over the dynamical fields, and $\bm\omega$ denotes their fluctuations. The brackets contain the first order expansion of the constraints $\delta S/\delta N^{a}$, which vanish by definition of $\bm n_{1}^{a}(\bm\omega)$. Hence we need only know the first order lapse and shift $\bm n_{1}^{a}(\bm\omega)$, which are still given by \eqref{ntnx}.

\subsection{Spatially flat gauge}
It is easiest to exhibit the slow roll suppression of the constrained cubic action in the spatially flat gauge \eqref{sfgauge}
\begin{equation}
\Lauxpert=0=\phipert 
\end{equation}
We expand the extended Liouville action \eqref{Lcovpre} plus scalar inflaton \eqref{SX} to third order in fluctuations, substitute the first order lapse and shift solutions \eqref{ntnx}, and discard $\Lauxpert$ and $\phipert$ terms. After a number of integrations by parts, one finds
\begin{multline}\label{S3x}
\SS_{3}= \int -\frac{\gamma}{4}\epsilon^{1/2} \bigl[ \infpert\infpertd^{2}+\infpert(\partial_{x}\infpert)^{2}\bigr]  -\psi\infpertd\partial_{x}\infpert \\
+\(  \frac{\gamma}{6}\epsilon^{3/2}\phib''
 + \frac{\gamma}{8}\epsilon^{1/2}a^{2}\VV_{,\infsub\infsub}(\infb) - \frac{1}{12}a^{2}\VV_{,\infsub\infsub\infsub}(\infb)
 \) \infpert^{3}
\end{multline}
This should be compared with the analogous expression in 4d (equation (3.8) of \cite{Maldacena:2002vr}). Upon wading through the notational differences,\footnote{The conversion between the notations employed here and in \cite{Maldacena:2002vr} is: $\phi$ there $\leftrightarrow$ $\infb$ here, $\varphi$ there $\leftrightarrow$ $\infpert$ here, $\rho$ there $\leftrightarrow$ $\phib$ here, $\zeta$ there $\leftrightarrow$ $\phipert$ here, $\partial_{i}\chi$ there $\leftrightarrow$ $\psi$ here.} accounting for the choice of time variable (coordinate time in \cite{Maldacena:2002vr}, conformal time here), and recalling the expression \eqref{SRvel} for $\epsilon$ in terms of the background velocities, one finds that most terms in equation (3.8) of \cite{Maldacena:2002vr} have precise analogs in \eqref{S3x}. The remaining terms are simply absent, because there are fewer distinct combinations of spatial derivatives in 1+1d.

Note that $\psi$ given by \eqref{psix} reduces in spatially flat gauge to
\begin{equation}\label{psisf}
\px\psi = - \frac{\gamma}{2}\epsilon\bigl(\infpert/\sqrt\epsilon\bigr)'
\end{equation}
Hence the entire first line of \eqref{S3x} is of order $\epsilon^{1/2}\infpert^{3}$. This common slow roll suppression is the advantage of working in spatially flat gauge. The second line of \eqref{S3x} is of higher order in slow roll; the relations \eqref{SRpot} imply that derivatives of the matter potential are suppressed as
\begin{equation}
\(\pd{}{\inflaton}\)^{n} \VV(\infb) \sim \epsilon^{n/2}
\end{equation}
Keeping only the first line of \eqref{S3x}, we substitute \eqref{psisf} to arrive at
\begin{equation}\label{S3sf}
\SS_{3} = -\frac{\gamma}{4} \int \epsilon^{1/2} \[ \infpert\infpert'^{2} +\infpert(\px\infpert)^{2} -2\infpert'\px\infpert\px^{-1}\infpert' \] 
-\frac{\gamma}{4} \int aH\epsilon^{1/2} \eta\, \infpert'\px\infpert\px^{-1}\infpert + \cdots
\end{equation}
Again, the last term may be discarded at leading order in slow roll. This is our final expression for the cubic action in spatially flat gauge. In the remainder of this section we clarify its order of slow roll suppression.

\subsubsection*{Translating between two gauges}
Thus far we have kept only the leading order slow roll terms, but it is not clear that these have the expected suppression by $\epsilon^{2}$. This is not surprising --- after all, the quadratic action \eqref{Meffact} is manifestly of order $\epsilon$ only when written in terms of $\zeta=\phipert$, the degree of freedom in constant inflaton gauge:
\begin{equation}\label{S2cf}
\SS_{2}^{\rm c.i.} = -\frac{2}{\gamma^{2}}\int \epsilon\, (\partial\phipert)^{2}
\end{equation}
In this section we have chosen spatially flat gauge (with degree of freedom $\infpert$), but it turns out to be useful to translate between $\infpert$ and $\phipert$ \cite{Maldacena:2002vr}. This is not the same as computing the cubic action in constant inflaton gauge, because \eqref{S3x} is an effective action where the constraints have been applied in spatially flat gauge. Neither are we performing a diffeomorphism directly on \eqref{S3sf}. Rather, the following is a change of variable arising from a comparison of the relative order in slow roll of $\infpert$ and $\phipert$. Of course, $\infpert$ and $\phipert$ are fluctuations which may have any profile whatsoever; the point is that a given profile of (say) $\infpert$ in spatially flat gauge induces a particular profile of $\phipert$ in constant inflaton gauge. We will work out this relation to second order in perturbations, which will be of use in the following section. Recall that constant inflaton gauge \eqref{cfgauge} is
\be\label{constx}
\infpert=0=\Lauxpert+\phipert
\ee
The condition $\Lauxpert+\phipert=0$ is trivially true in spatially flat gauge ($\Lauxpert=0=\phipert$), and according to the gauge transformations \eqref{fieldtransf} is preserved under time reparametrizations $\xi^{\ct}$. To access the gauge slice $\infpert=0$ beginning from spatially flat gauge, one must make a diffeomorphism $(\xi^{\ct},\xi^{x})$ satisfying
\begin{equation}\label{xiteq}
\infb(x^{a}+\xi^{a})+\infpert\sfg(x^{a}+\xi^{a})=\infb(x^{a})
\end{equation}
where $\infpert\sfg(x^{a})$ is the profile of $\infpert$ in spatially flat gauge. We wish to keep $\xi^{x}=0$, and solve \eqref{xiteq} for $\xi^{\ct}$ up to second order in $\infpert\sfg$. The Taylor expansion of $\infpert\sfg(x^{a}+\xi^{a})$, which was omitted from the gauge transformation \eqref{infgaugetrans}, must be included. The solution to \eqref{xiteq} is then
\begin{equation}\label{xit2}
\infb' \xi^{\ct}= - \infpert\sfg + \half \bigl( \infpert\sfg^{2}/\infb' \bigr)'
\end{equation}
This is the diffeomorphism which takes us from spatially flat gauge to constant inflaton gauge. The profile of $\phipert$ in this new gauge is
\begin{equation}\label{chi2}
\begin{aligned}
\phipert\cfg &=\phib'\xi^{\ct}+\half\phib''(\xi^{\ct})^{2} \\
&= - \frac{\phib'}{\infb'}\infpert\sfg + \frac{1}{2\infb'}\bigl( \phib'\infpert\sfg^{2}/\infb' \bigr)' 
\end{aligned}
\end{equation}
In terms of the slow roll parameters,
\begin{equation}\label{phi2x}
\phipert\cfg =  - \frac{1}{z}\,\infpert\sfg 
+ \frac{1}{2aHz}\bigl( \infpert\sfg^{2}/z \bigr) '
\end{equation}
The second order inverse of this relation is
\begin{equation}\label{x2phi}
\infpert\sfg = -z\phipert\cfg + \frac{1}{2aH}\bigl( z\phipert\cfg^{2} \bigr) '
\end{equation}
Recall from \eqref{zdef} that $z\propto\epsilon^{1/2}$. The first order relation $\infpert\sfg\sim\epsilon^{1/2}\phipert\cfg$ renders the cubic action \eqref{S3sf} of order $\epsilon^{2}$ in terms of $\phipert\cfg$. To cover all the bases, one must also account for the mixing of quadratic and cubic actions arising from the second order term in \eqref{x2phi}. This does not spoil the $\epsilon^{2}$ suppression of the cubic action. An explicit demonstration of this is postponed to the following section, where a similar second order field redefinition arises.

\subsection{Constant inflaton gauge}
We now rederive the cubic action in constant inflaton gauge \eqref{cfgauge}, where the degree of freedom is $\zeta=\phipert$, the perturbation of the scale factor. This gauge will be advantageous for our computation of the three-point function in section \ref{3ptsec}, because once $\phipert$ leaves the horizon it is known to be constant. We can just evaluate the three-point function of $\phipert$ up to the time of horizon crossing, avoiding the transition from inflation to reheating, where the slow roll parameters are no longer small.

Again expanding the action \eqref{Lcovpre} plus \eqref{SX} to cubic order, substituting the first order lapse and shift \eqref{ntnx}, imposing the gauge conditions, and integrating by parts, we get
\begin{equation}\label{S3cf}
\SS_{3} = \frac{2}{\gamma^{2}}\int 
 \epsilon\, \Bigl\{ \phipert\phipert'^{2} + \phipert(\px\phipert)^{2} + \frac{1}{aH}\phipert'(\partial\phipert)^{2}
- 2\psi\phipert'\px\phipert \Bigr\}
\end{equation}
The analogous expression in 4d is equation (3.9) of \cite{Maldacena:2002vr}. In constant inflaton gauge, equation \eqref{psix} for $\psi$ reduces to
\begin{equation}\label{psicf}
\px\psi = \epsilon\phipert'
\end{equation}
At this point, only the $\psi$ term in \eqref{S3cf} is manifestly suppressed by $\epsilon^{2}$ --- the other terms are suppressed only by $\epsilon$. Substituting $\phipert\cfg=-\infpert\sfg/z+\cdots$ \eqref{phi2x} would make matters worse, but there is a related field redefinition which happens to be useful. We know from the previous section that the relevant terms in the cubic action \eqref{S3sf} in spatially flat gauge have a common order of slow roll suppression. It turns out that the action \eqref{S3cf} adopts a similar form in terms of a new variable $\phin$ satisfying \cite{Maldacena:2002vr}
\begin{equation}\label{phindef}
\phin\cfg = -\infpert\sfg /z
\end{equation}
exactly, to all orders in perturbations. Compare this with our previous statement \eqref{phi2x} relating variables in the two gauges. Using \eqref{phindef} to eliminate $\infpert\sfg$ from \eqref{phi2x}, the relation between $\phipert$ and $\phin$ to second order is
\begin{equation}\label{phi2phin}
\phipert = \phin + \frac{\eta}{4}\,\phin^{2} + \frac{1}{aH}\phin\phin'
\end{equation}
This may be compared with equation (3.10) of \cite{Maldacena:2002vr}. At first order $\phipert=\phin$, so the change of variable $\phipert\to\phin$ does not alter the quadratic action \eqref{S2cf}. It does however give rise to additional third order terms:
\begin{equation}
\SS_{2}[\phipert] = \SS_{2}[\phin] + 
\frac{4}{\gamma^{2}}\int \epsilon\, \Bigl(  \frac{\eta}{4}\,\phin^{2} + \frac{1}{aH}\phin\phin' \Bigr) \Bigl( \partial^{2}\phin - aH\eta\phin' \Bigr)
\end{equation}
where the last bracket is proportional to the equation of motion $\delta\SS_{2}/\delta\phin$. Adding these terms to \eqref{S3cf} and integrating by parts, the $O(\epsilon)$ terms in \eqref{S3cf} are canceled, while new $O(\epsilon^{2})$ terms are generated. The result is
\begin{equation}\label{S3cffinal}
\SS_{3} = \frac{2}{\gamma^{2}}\int \epsilon^{2}\Bigl( \phin\phin'^{2} + \phin(\px\phin)^{2} -2\phin'\px\phin\px^{-1}\phin' \Bigr)  - \frac{1}{3\gamma^{2}}\int (\epsilon\eta')' \,\phin^{3}
\end{equation}
This cubic action is correct to all orders in slow roll. It is of a very similar form to the spatially flat action \eqref{S3sf} --- in fact, the change of variable \eqref{phindef} shows they are identical up to the non-derivative terms (which are further slow roll suppressed). The point of using the variable $\varn$ is that each order in the perturbative action is suppressed by an additional factor of $\epsilon$. The only disadvantage is that $\varn$ is not constant outside the horizon. 

It is worth noting that the action \eqref{S3cffinal} could be simplified even further by another change of variable \cite{Seery:2005wm}. The idea is to integrate by parts, beginning with $\px^{-1}\phin'$ to get $\px^{-2}\phin'$. Whenever second derivatives $\px^{2}\phin$ or $\phin''$ arise, they are eliminated in favor of $\delta\SS_{2}/\delta\phin$, plus other terms occurring in the equation of motion. Continuing in this way, the leading cubic action may be expressed as a single vertex plus terms proportional to $\delta\SS_{2}/\delta\phin$, which are then removed by a field redefinition. Rather than make this additional field redefinition, we choose to work with the action \eqref{S3cffinal}.

\section{Three-point function}\label{3ptsec}
We now proceed with the computation of the tree-level three-point function for the curvature perturbation $\zeta=\phipert$. The constancy of $\phipert$ outside the horizon makes it suitable for computing observable signatures at the end of inflation. Following \cite{Maldacena:2002vr}, the strategy is to begin with the variable $\phin$, in terms of which there is a perturbation expansion governed by the slow roll parameter $\epsilon$. The three-point function of $\varn$ is evaluated in section \ref{phibarsec} using the leading order terms in the cubic action \eqref{S3cffinal}. After the modes in question exit the horizon, we switch to the variable $\var$, which is then constant. This allows us to infer the three-point function for all subsequent times. The change of variable $\varn\to\var$ induces correction terms in the three-point function, which are given in section \ref{redefsec}. Finally, in section \ref{consistencysec} we discuss symmetry constraints on the three-point function.

Working in momentum space, the Fourier transform of $\var(\ct,\bx)$ is written in terms of the mode functions defined in \eqref{vexp} as
\begin{equation}
\var(\ct,\bk) =  a(\bk)\var_{k}(\ct) + a^{\dagger}(-\bk)\var_{k}^{*}(\ct) 
\end{equation}
The spatial momentum $\bk$ in 1+1d is just a real number of either sign; the magnitude of $\bk$ is denoted by $k$. In the interaction picture, the contraction of Fourier components is
\begin{equation}\label{cont}
\vacl \var(\ct,\bk)\var(\ct',\bp) \vacr = \var_{k}(\ct) \var_{p}^{*}(\ct') (2\pi)\delta(\bk+\bp)
\end{equation}
The Bunch-Davies mode function $\var_{k}(\ct)$ in the slow roll approximation is equal to the Hankel function \eqref{vksol}, scaled by $-1/z$. To evaluate the three-point function to lowest order in slow roll, we need only the lowest order expression for the mode function, equation \eqref{zetak2d}. The three-point function to be computed is
\begin{equation}
\lim_{\ct\to 0^{-}} \big\langle \var(\ct,\bp_{1})\var(\ct,\bp_{2})\var(\ct,\bp_{3}) \big\rangle_{\rm tree}
\end{equation}
The limit $\ct\to 0^{-}$ takes us to the end of inflation; in the following equations we simply specialize to $\ct=0$.

\subsection[Three-point function of phi-bar]{Three-point function of $\varn$}\label{phibarsec}
The three-point function of $\varn$ is evaluated using the leading terms in the cubic action \eqref{S3cffinal}. In the in-in formalism, it takes the form of an expectation value along a closed time contour extending from $\ct=-\infty$ to $\ct=0$, and back again \cite{LimNotes}:
\begin{equation}\label{3pt}
\begin{aligned}
\big\langle \varn(0,\bp_{1}) & \varn(0,\bp_{2})\varn(0,\bp_{3}) \big\rangle_{\rm tree} \\
&=
\Ivacl U_{\rm int}(0,-\infty)^{-1} \varn(0,\bp_{1})\varn(0,\bp_{2})\varn(0,\bp_{3}) U_{\rm int}(0,-\infty) \Ivacr_{\rm tree}
\\
&=\lim_{T\to-\infty(1-i\varepsilon)} 2\,\mathrm{Re}\, \vacl\, \varn(0,\bp_{1})\varn(0,\bp_{2})\varn(0,\bp_{3}) \Bigl[ -i\int_{T}^{0}d\ct H_{3}(\ct) \Bigr] \vacr + \cdots 
\end{aligned}
\end{equation}
All fields are in the interaction picture, with time dependence determined by the free Hamiltonian. The remaining time dependence is carried by the interacting evolution operator $U_{\rm int}(\ct_{\rm f},\ct_{\rm i})$, which is treated perturbatively. The free vacuum $|0\rangle$ is not an eigenstate of the interacting Hamiltonian, but rather an infinite sum of eigenstates. Of these, the lowest eigenvalue belongs to the interacting vacuum $|\Omega\rangle$, which is projected out by an infinitesimal rotation of the time contour, moving the lower terminal from $-\infty$ to $-\infty(1-i\varepsilon)$. Here $\varepsilon$ is an infinitesimal constant, not to be confused with the slow roll parameter $\epsilon$.

At cubic order in $\varn$ the interacting Hamiltonian is simply $H_{3}(\pi,\varn)=-L_{3}(\varn',\varn)$, for the following reason. The conjugate momentum is of the form $\pi(\varn)=\epsilon\varn'+\mathcal{O}(\epsilon^{2}\varn^{2})$, where the nonlinear terms come from the interacting Lagrangian. The Hamiltonian is constructed as the Legendre transform
\begin{equation}\label{Legtrans}
H = \pi\,\varn'(\pi)-L_{2}-L_{3}+\cdots
\end{equation}
The cubic contributions to the first two terms cancel, and the remaining quadratic terms are identified with the free Hamiltonian. The only other cubic contribution to \eqref{Legtrans} comes from $L_{3}$, which gives $H_{3}(\pi,\varn)=-L_{3}(\varn',\varn)+\mathcal{O}(\epsilon^{3}\varn^{4})$. Hence the cubic Hamiltonian at lowest order in slow roll is
\begin{equation}\label{H3}
H_{3}(\ct) = - \frac{2}{\gamma^{2}}\,\epsilon(\ct)^{2} \int dx\, \Bigl( \phin\phin'^{2} + \phin(\px\phin)^{2} -2\phin'\px\phin\px^{-1}\phin' \Bigr) 
\end{equation}

The contribution of the first term in \eqref{H3} (which we denote by $H_{3}^{(1)}$) to the three-point function \eqref{3pt} will now be evaluated explicitly. Going to momentum space and using the contraction \eqref{cont}, one finds the expectation value
\begin{multline}\label{3ptstep}
\vacl \varn(0,\bp_{1})\varn(0,\bp_{2})\varn(0,\bp_{3}) H_{3}^{(1)}(\ct) \vacr 
= -\frac{8\pi}{\gamma^{2}}\, \epsilon(\ct)^{2}\, 
\var_{p_{1}}\var_{p_{2}}\var_{p_{3}}(0)\, \var_{p_{1}}^{*}\var_{p_{2}}^{*}\!\!'\, \var_{p_{3}}^{*}\!\!'(\ct)\, \delta\(\sum\bp_{i}\)   \\
+ (1\leftrightarrow2) + (1\leftrightarrow3)
\end{multline}
Note that the mode functions $\var_{k}$ and $\varn_{k}$ are identical, because they have the same quadratic action ($\var=\varn$ at lowest order). The momentum-conserving delta function will be implicit in the following expressions for correlation functions. The contribution of $H_{3}^{(1)}$ to the three-point function \eqref{3pt} is then
\begin{equation}\label{3ptstep2}
\big\langle \varn(0,\bp_{1})\varn(0,\bp_{2})\varn(0,\bp_{3}) \big\rangle ' _{\rm tree} \supset
\frac{16\pi}{\gamma^{2}}\, \mathrm{Re}\[ i\, \var_{p_{1}}\var_{p_{2}}\var_{p_{3}}(0)  \int_{-\infty(1-i\varepsilon)}^{0}d\ct\, \epsilon(\ct)^{2} \var_{p_{1}}^{*}\var_{p_{2}}^{*}\!\!'\, \var_{p_{3}}^{*}\!\!'(\ct) \]
\end{equation}
This should be evaluated to lowest order in slow roll, because we are already neglecting higher terms by restricting to the cubic Hamiltonian. We use the mode functions $\phipert_{k}=\zeta_{k}$ given in \eqref{zetak2d}:
\begin{equation}
\var_{k}(\ct) = - \frac{1}{z(\ct)} \frac{e^{-ik\ct}}{\sqrt{2k}}
\end{equation}
Up to the moment of horizon crossing, this is an excellent approximation to the full Hankel function. At this level of approximation, $z(\ct)\propto\epsilon(\ct)^{1/2}$ is a constant, which we take to be $z(\ct_{*})$, where $\ct_{*}$ is the moment of horizon crossing for the shortest wavelength mode (the last mode to leave the horizon). Strictly speaking, the upper terminal of the time integral in \eqref{3ptstep2} should be equal to $\ct_{*}$, but the additional contribution from the interval $\ct\in(\ct_{*},0)$ is dwarfed by the contribution before horizon crossing, $\ct<\ct_{*}$. Hence we can approximate the time integral by
\begin{equation}
\int_{-\infty(1-i\varepsilon)}^{0}d\ct\, e^{iP\ct} = \frac{1}{iP}
\end{equation}
where $P$ denotes the sum of the energies (magnitudes) carried by the three momenta,
\begin{equation}
P=\sum p_{i}
\end{equation}
The contribution of $H_{3}^{(1)}$ to the three-point function \eqref{3pt} is then
\begin{equation}\label{H31contrib}
\big\langle \varn(0,\bp_{1})\varn(0,\bp_{2})\varn(0,\bp_{3}) \big\rangle  _{\rm tree} \supset
-\frac{\pi\gamma^{4}}{32\epsilon_{*}}\, \frac{1}{p_{1}p_{2}p_{3}}\, \frac{1}{2P} \sum_{i\neq j}p_{i}p_{j}
\end{equation}
Recall that the sum of the momenta, $\sum\bp_{i}$, is constrained to vanish by an implicit momentum-conserving delta function.

The remaining terms in the Hamiltonian \eqref{H3} make similar contributions, differing only by the form of the final momentum sum in \eqref{H31contrib}. The contribution of the second term in \eqref{H3} is given by replacing in \eqref{H31contrib}
\begin{equation}
\sum_{i\neq j}p_{i}p_{j} \to \sum_{i\neq j}\bp_{i}\bp_{j} = -\sum_{i}p_{i}^{2} 
\end{equation}
where momentum conservation was used. The contribution of the final term in \eqref{H3} is given by replacing
\begin{equation}\label{sum3}
\sum_{i\neq j}p_{i}p_{j} \to - 2\Bigl( p_{1}\bp_{2}\, \mathrm{sign}(\bp_{3}) +\text{perms} \Bigr) 
\end{equation}
where all six permutations of the momenta are to be included. Again using momentum conservation, this is written in terms of the magnitudes as
\begin{equation}
- 2\Bigl( p_{1}\bp_{2}\, \mathrm{sign}(\bp_{3}) +\text{perms} \Bigr)
= \frac{p_{1}^{2}}{p_{2}p_{3}}\(p_{1}^{2}-p_{2}^{2}-p_{3}^{2}\) + (1\lr2) + (1\lr3)
\end{equation}
The complete three-point function of $\varn$ is then
\begin{equation}\label{varn3ptfinal}
\big\langle \varn(0,\bp_{1})\varn(0,\bp_{2})\varn(0,\bp_{3}) \big\rangle  _{\rm tree} =
-\frac{\pi\gamma^{4}}{64\epsilon_{*}}\,  F(\bp_{1},\bp_{2},\bp_{3})
\end{equation}
where the shape function is
\begin{equation}\label{shape1}
F(\bp_{1},\bp_{2},\bp_{3}) = \frac{1}{p_{1}p_{2}p_{3}}\,\frac{1}{P} \biggl[ -\sum_{i}p_{i}^{2}  + \sum_{i\neq j}p_{i}p_{j} 
+ \frac{p_{1}^{2}}{p_{2}p_{3}}\(p_{1}^{2}-p_{2}^{2}-p_{3}^{2}\) + (1\lr2) + (1\lr3)
 \biggr]
\end{equation}
This may be compared with the local shape function generated by single field inflation in 4d, given explicitly in equation (223) of \cite{LimNotes}. The first two terms in \eqref{shape1} are 2d analogs of the first two terms proportional to $\epsilon$ in \cite{LimNotes}. The final term is different --- at a superficial level, this stems from the appearance in \eqref{sum3} of $\mathrm{sign}(\bp_{i})$. Instead of further massaging \eqref{shape1} to more closely resemble the 4d result, we prefer to express it in a very compact form, taking advantage of the simplified kinematics of spatial momenta in 1+1d. Imposing momentum conservation explicitly and assuming various kinematic configurations, one finds in every possible case that
\begin{equation}
F(\bp_{1},\bp_{2},\bp_{3})  = F(\bp_{1},\bp_{2},-\bp_{1}-\bp_{2}) = 16/P^{2}
\end{equation}
Hence the three-point function of $\varn$ has a completely flat shape, depending only on the total energy carried by the three momenta. The change of variable $\varn\to\var$ will introduce dependence on the shape of the spatial momentum ``triangle.''

\subsection{Field redefinition}\label{redefsec}

Once the modes in question are outside the horizon we switch to the variable $\var$, which is then constant to all orders in slow roll (and all orders in perturbations). The inverse of the second order relation \eqref{phi2phin} between $\phipert$ and $\phin$ is
\begin{equation}\label{phin2phi}
\varn = \var - \frac{\eta}{4}\var^{2} - \frac{1}{H}\var\dot\var
\end{equation}
Recall that a dot denotes the coordinate time derivative $\partial_{t}$. The final term in \eqref{phin2phi} vanishes outside the horizon, so that $\varn$ has only a weak time dependence inherited from $\eta$:
\begin{equation}
\dot\phin = -\frac{1}{4}\dot\eta\,\phipert^{2} = -\frac{1}{4}H\eta\kappa\, \phipert^{2}
\qquad\text{(superhorizon)}
\end{equation}
where $\kappa$ was defined in \eqref{SRparams}. Returning to the original relation \eqref{phi2phin} between $\phipert$ and $\phin$, the $\dot\varn$ term is of higher order in slow roll and may be dropped, leaving
\begin{equation}\label{phi2phinsuper}
\phipert = \phin + \frac{\eta}{4}\,\phin^{2}
\qquad \text{(superhorizon)}
\end{equation}
This leads to the following relation between three-point functions in momentum space:
\begin{multline}\label{3ptcorrec1}
\big\langle \var(\bp_{1})\var(\bp_{2})\var(\bp_{3}) \big\rangle 
= \big\langle \varn(\bp_{1})\varn(\bp_{2})\varn(\bp_{3}) \big\rangle \\
+ \frac{\eta}{4} \int\frac{d\bk}{2\pi} \Bigl[ \big\langle\varn(\bp_{1}-\bk)\var(\bp_{2})\big\rangle \big\langle\varn(\bk)\var(\bp_{3})\big\rangle +\text{perms} \Bigr] + \cdots
\end{multline}
The time arguments of the fields have been suppressed for brevity, but are of course coincident at some time after horizon crossing. The ellipsis represents terms of higher order in slow roll, corresponding to more than one insertion of \eqref{phi2phinsuper}. To evaluate the corrections \eqref{3ptcorrec1} to lowest order in slow roll, note that $\langle\varn\var\rangle=\langle\var\var\rangle$ in the free theory, because $\var$ and $\varn$ share the same quadratic action. Then, for instance, the first correction term on the right hand side of \eqref{3ptcorrec1} is
\begin{equation}
\begin{aligned}
\int\frac{d\bk}{2\pi}  \big\langle\varn(\bp_{1}-\bk)\var(\bp_{2})\big\rangle \big\langle\varn(\bk)\var(\bp_{3})\big\rangle
 &= \delta\(\sum \bp_{i}\) (2\pi) |\var_{p_{2}}|^{2} |\var_{p_{3}}|^{2} (\ct_{*}) \\
& =  \delta\(\sum \bp_{i}\) \frac{\pi\gamma^{4}}{32\epsilon_{*}^{2}}\, \frac{1}{p_{2}p_{3}}
\end{aligned}
\end{equation}
The total relation \eqref{3ptcorrec1} between three-point functions becomes
\begin{equation}\label{3ptcorrec}
\big\langle \var(\bp_{1})\var(\bp_{2})\var(\bp_{3}) \big\rangle 
= \big\langle \varn(\bp_{1})\varn(\bp_{2})\varn(\bp_{3}) \big\rangle 
-\frac{\pi\gamma^{4}}{64}\, \frac{\eta_{*}}{\epsilon_{*}^{2}}\, \frac{P}{p_{1}p_{2}p_{3}}
\end{equation}
This is the precise analog of the $\eta$-dependent term in the 4d shape function, equation (223) of \cite{LimNotes}. Substituting our result \eqref{varn3ptfinal} for $\langle\varn\varn\varn\rangle$, we find the three-point function for $\zeta=\phipert$ at the end of inflation is
\begin{equation}\label{phi3pt}
\begin{aligned}
\big\langle \var(0,\bp_{1})\var(0,\bp_{2})\var(0,\bp_{3}) \big\rangle _{\rm tree} &= 
-\frac{\pi\gamma^{4}}{64\epsilon_{*}}\biggl(  F(\bp_{1},\bp_{2},\bp_{3}) 
+ \frac{\eta_{*}}{\epsilon_{*}}\, \frac{P}{p_{1}p_{2}p_{3}}   \biggr)    \\
&= -\frac{\pi\gamma^{4}}{64\epsilon_{*}}\biggl(  \frac{16}{P^{2}} 
+ \frac{\eta_{*}}{\epsilon_{*}}\, \frac{P}{p_{1}p_{2}p_{3}}   \biggr) 
\end{aligned}
\end{equation}
This is a local shape, which diverges in the squeezed limit (where one of the momenta tends to zero). Hence the greatest non-Gaussianity is associated to squeezed configurations. For a given value of the total energy $P$, the minimum magnitude of \eqref{phi3pt} (least non-Gaussianity) occurs when two of the momenta are identical, for instance $\bp_{1}=\bp_{2}=-\bp_{3}/2$. This corresponds to a folded isosceles triangle, which is the closest one can get to an equilateral triangle in one spatial dimension.

\subsection{Consistency relation}\label{consistencysec}

\subsubsection*{Residual diffeomorphisms and global shift symmetry}
Recently much attention has been devoted to Ward identities for cosmological correlation functions \cite{Hinterbichler:2013dpa,Berezhiani:2013ewa,Pimentel:2013gza,Creminelli:2012qr}. These identities may be understood to arise from residual diffeomorphism symmetry. Our choice of constant inflaton gauge \eqref{cfgauge} completely fixes diffeomorphisms which vanish at spatial infinity. Let us define ``physical'' metric and matter fluctuations as those vanishing at spatial infinity. Residual gauge symmetries are the subset of diffeomorphisms not vanishing at infinity, which preserve the gauge choice and the boundary conditions (physicality) of fluctuations. In 4d, the residual symmetries in constant inflaton gauge were classified in \cite{Hinterbichler:2013dpa}. Following the same arguments in 2d, we begin by looking for diffeomorphisms which preserve the gauge choice \eqref{cfgauge}:
\begin{equation}\label{cfgauge2}
\infpert = 0 = \Lauxpert + \phipert
\end{equation}
Consulting the gauge transformations \eqref{fieldtransf} and \eqref{infgaugetrans}, we see that a time reparametrization $\xi^{\ct}$ breaks the inflaton gauge condition, due to the time-dependent inflaton background. The gauge transformation of $\Lauxpert+\phipert$ is
\begin{equation}\label{chiplusphitrans}
(\Lauxpert+\phipert) \to (\Lauxpert+\phipert) + \px^{2}\xi
\end{equation}
Recall that the spatial reparametrization \eqref{coordtransf4d} is $x\to x+\px\xi$, so there is no meaning to $x$-independent reparametrizations $\xi$. The gauge condition $\Lauxpert+\phipert=0$ is preserved by spatial translations $\xi\sim x$, but these don't give rise to any Ward identities, as our correlation functions already conserve spatial momentum. In 3+1d there are transverse spatial reparametrizations, which preserve the gauge choice, but there is no room for these in 1+1d. This leaves an arbitrary time dependence of $\xi$ as the only freedom. In summary, the gauge-preserving diffeomorphisms are of the form
\begin{equation}\label{gaugepreserving}
\xi^{\ct}(\ct,x)=0~,\qquad \xi(\ct,x)=x\, \overline{\xi}(\ct)
\end{equation}

Now let's consider the physicality condition, that the residual gauge symmetry must preserve the vanishing of metric fluctuations at spatial infinity. Diffeomorphisms of the form \eqref{gaugepreserving} automatically preserve the physicality of $\Lauxpert$, $\phipert$, and $\infpert$. We still have to check the physicality of the lapse and shift fluctuations $\ntpert$ and $\nxpert$, which were adjusted to solve the perturbed constraints \eqref{constraintpert}. The constraints are gauge invariant (by the closure of the constraint algebra), but in solving them we must also choose a boundary condition: the vanishing at spatial infinity of the function $\psi$, which in constant inflaton gauge is given by \eqref{psicf}. The variation of $\psi$ under a gauge-preserving diffeomorphism \eqref{gaugepreserving} is
\begin{equation}
\psi \to \psi + \epsilon\, \overline{\xi}\, '
\end{equation}
Hence $\overline{\xi}\, '$ must vanish at spatial infinity; but $\overline{\xi}(\ct)$ is $x$-independent, so it must be a constant. The only diffeomorphism which preserves constant inflaton gauge and the physicality of fluctuations is a spatial translation, and there are no Ward identities!

This cannot be the whole story, however, because the three-point function \eqref{phi3pt} does in fact obey Maldacena's consistency relation\footnote{The sign of this relation differs in some parts of the literature, due to a different convention for the sign of the scalar curvature perturbation.} \cite{Maldacena:2002vr}
\begin{equation}\label{Maldacena2d}
\lim_{q\to0}\, \frac{\langle\var(\bq)\var(\bp)\var(-\bp-\bq)\rangle}{P_{\var}(\bq)} = 2\pi(n_{s}-1)_{*}\, P_{\var}(\bp)
\end{equation}
The sole contribution to the left side in the limit comes from the second term in \eqref{phi3pt}, which translates between the $\phipert$ and $\phin$ three-point functions. This term is proportional to $\eta$, as is the spectral tilt \eqref{tilt}. Maldacena's consistency relation in 4d is known to arise from a residual gauge symmetry under spatial dilations (see for example \cite{Hinterbichler:2013dpa}). In 2d a spatial dilation is given by $\xi\sim x^{2}$, under which $\Lauxpert+\phipert$ shifts by a constant, according to \eqref{chiplusphitrans}. The key point is that the gauge condition $\Lauxpert+\phipert=0$ may be restored using the global shift symmetry of $\Laux$ discussed in section \ref{Liouvillesec}, and that this is only possible because $\Lauxpert+\phipert$ shifts by a constant under spatial dilations. The shift symmetry of $\Laux$ arises from the existence of the topological Euler characteristic, and is special to 2d. It is a physical symmetry rather than a gauge symmetry. Spatial diffeomorphisms more complicated than a dilation disturb the gauge condition in ways not related by any symmetry.

A similar shuffling of gauge transformations occurs in 4d, where there is an ambiguity in the parametrization \eqref{ds4d} of scalar perturbations of the space-space metric: the contribution proportional to $\delta_{ij}$ may be shuffled between $\Psi$ and $E$. The ambiguity is resolved by requiring $E$ to be harmonic: $\vec\partial^{2}E=0$. Part of the constant inflaton gauge choice is the condition $E=0$, although strictly speaking this should be $E_{,ij}=0$, because only $E_{,ij}$ appears in the metric \eqref{ds4d}. A typical spatial diffeomorphism $x^{i}\to x^{i}+\partial^{i}\xi$, which sends $E\to E+\xi$, will break the gauge condition by a term $\xi_{,ij}$. If this is proportional to $\delta_{ij}$ then it may be absorbed in $\Psi$, so residual diffeomorphisms are required to preserve $E_{,ij}=0$ only up to such terms. Indeed, the spatial dilation $\xi=(\lambda/2) x_{i}x^{i}$ --- which is the source of Maldacena's consistency relation --- falls into this category, inducing $E_{,ij}=\lambda\delta_{ij}$. In 2d there is no ambiguity in the parametrization of space-space metric fluctuations, but when $\px^{2}\xi$ is constant, the global shift symmetry of extended Liouville theory plays a similar role.

For the sake of completeness, let us mention how the consistency relation \eqref{Maldacena2d} associated to spatial dilations is arrived at in 2d. It is the leading term in a master consistency relation, similar to those motivated in 4d by examining gauge transformations of the path integral for the quantum effective action \cite{Berezhiani:2013ewa}. Following an identical argument in 2d, one finds the master consistency relation
\begin{equation}\label{master2d}
\frac{1}{2\pi}\, \frac{\langle\var(\bq)\var(\bp)\var(-\bp-\bq)\rangle}{P_{\var}(\bq)} = -P_{\var}(\bp+\bq) - \bp \[\frac{P_{\var}(\bp+\bq)-P_{\var}(\bp)}{\bq}\]
\end{equation}
Recall that $\bp$ and $\bq$ are just real numbers in 1+1d. The analogous expression in 3+1d (equation (4.2) of \cite{Berezhiani:2013ewa}) is proportional to the momentum $q^{i}$ with free spatial index. Maldacena's relation \eqref{Maldacena2d} is the $q^{0}$ term in the Taylor expansion of \eqref{master2d} about $\bq=0$, and is associated to spatial dilations. Higher terms in the Taylor expansion are associated to more complicated spatial diffeomorphisms. It was demonstrated above that none of these preserve the gauge condition in 2d, so they are not symmetries of the three-point funcion \eqref{phi3pt}. If it weren't for this, \eqref{master2d} would completely determine the three-point function to all orders in $q$. This is in contrast to 3+1d, where there is an infinite hierarchy of consistency relations but still room for model dependence, due to the existence of transverse spatial polarizations.

\subsubsection*{Global Weyl symmetry}

Finally, the shift symmetry of $\Laux$ may also be thought of as a global Weyl symmetry in constant inflaton gauge, which is related to the familiar local Weyl symmetry of regular Liouville theory. To see this, begin by substituting $g_{ab}=e^{2\phican}\ghat_{ab}$ in the action \eqref{Lcovpre}, upon which it adopts the form \eqref{Sgravhat}. No gauge choice has yet been assumed if $\ghat$ is allowed to fluctuate. In backgrounds satisfying $\Lauxb'+\phib'=0$, the constant inflaton gauge condition $\Lauxpert+\phipert=0$ makes the full fields (background plus perturbation) satisfy $\Laux+\phican=C$, a constant. This constant contributes to the action only as a multiple of the Euler characteristic \eqref{Euler}, so it may be neglected. We may then impose the gauge condition $\Laux+\phican=0$ (ie. $\Theta=0$) on the action \eqref{Sgravhat}, reducing it to
\be\label{Sci}
\SS\cfg = \frac{2}{\gamma^2} \int \sqrt{-\ghat}\, \[(\hat\nabla\phican)^2 + \hat\R\phican
-{\Lambda} e^{2\phican} \]
\ee
This is precisely the action for timelike Liouville theory \cite{Martinec:2014uva}, but with a different interpretation. The action \eqref{Sci} is usually studied in conformal gauge, where it describes the dynamics of the Liouville field $\phican$ in a fixed reference metric $\ghat$. It is well known to possess a symmetry under local Weyl transformations \cite{Ginsparg:1993is}
\begin{equation}\label{Weyl}
\ghat_{ab}\to e^{2\omega}\ghat_{ab}~, \qquad
\phican \to \phican - \omega
\end{equation}
where $\omega$ is an arbitrary function of space and time. This transformation may appear trivial (as it leaves $g_{ab}=e^{2\phican}\ghat_{ab}$ invariant), but it does in fact change the gauge choice. In conformal gauge we exhaust the local diffeomorphism symmetry to fix $\ghat$, and there is a family of conformal gauges where $\ghat$ has been fixed to different values. The Weyl transformation \eqref{Weyl} moves along a slice in this family. This is not a residual gauge symmetry, and there is no a priori reason to expect the gauge-fixed action \eqref{Sci} to be invariant. In fact, the action changes by
\begin{equation}\label{STLvar}
\SS\cfg  \to \SS\cfg - \frac{2}{\gamma^2} \int \sqrt{-\ghat}\, \hat R\, \omega
\end{equation}
The change depends only on $\ghat$, which is fixed in conformal gauge. The $\phican$ equation of motion is unaltered, so the local Weyl transformation \eqref{Weyl} is a symmetry of Liouville theory in conformal gauge.

In the present context, we are instead studying \eqref{Sci} in constant inflaton gauge. Both $\phican$ and $\ghat$ are fluctuating, as the gauge freedom has been used to eliminate $\Lauxpert$ and fix $\infpert=0$. The variation \eqref{STLvar} of the gauge-fixed action depends on $\ghat$, and in general will alter its equation of motion. If $\omega$ is a constant, however, \eqref{STLvar} is just a multiple of the Euler characteristic. Hence the action \eqref{Sci} possesses only a global Weyl symmetry in this context. The consequences of the global Weyl transformation for cosmological perturbation theory are made clear if it is packaged as
\begin{equation}\label{Weylglobal}
\phib\to\phib+\omega~,\qquad \phipert\to\phipert-\omega
\end{equation}
This doesn't interfere with the metric parametrization \eqref{metparam}. Again, it is not a trivial transformation as the gauge has already been fixed. The slow roll expansion of the constrained action depends on $\phib$ only through the slow roll parameters, which are invariant under \eqref{Weylglobal} (the transformation of $\phib$ is just a constant rescaling of the scale factor). Hence the constant shift of $\phipert$ is a physical symmetry of the correlation functions.

\section{Conclusion}
In this paper we have applied the framework of cosmological perturbation theory to single field inflation in a generally covariant extension of Liouville theory. Our main results are the unified power spectrum \eqref{P2}, and the three-point correlation function of scalar curvature perturbations, equation \eqref{phi3pt}. We have investigated symmetry constraints on the three-point function and explained how Madacena's consistency condition \eqref{Maldacena2d} arises, despite the absence of residual diffeomorphisms. An important subresult is the proof of freeze out to all orders in perturbation theory, given in section \ref{constancysec}.

This work completes the construction of a simplified 2d laboratory for the scalar sector of cosmological perturbations in 4d Einstein gravity. We now possess a precise model of the scalar sector embedded in a renormalizable theory. A future publication will explore renormalized perturbations in cosmological gauges, addressing various aspects including the quantum-corrected constraint algebra, and the classification of physical states via BRST cohomology.

\section*{Acknowledgments}
It is a pleasure to thank Emil Martinec for suggesting this topic, for discussions, and detailed comments on a draft manuscript. I am also grateful to Wayne Hu, Austin Joyce, and Guilherme Pimentel for discussions. This work was supported in part by DOE grant DE-FG02-13ER41958, and by the US Department of State through a Fulbright Science and Technology Award.

\singlespacing

\bibliographystyle{JHEP}
\bibliography{nGbib}

\end{document}